\documentclass[twocolumn,aps,superscriptaddress,showpacs,floatfix,prc]{revtex4-1}
\usepackage{mathrsfs}
\usepackage{amssymb}
\usepackage{amsmath}
\usepackage{graphicx}
\usepackage[normalem]{ulem}
\usepackage[dvips]{color}
\usepackage{bm}
\usepackage{longtable}\usepackage{slashed}

\usepackage{newtxtext}
\usepackage{enumitem}
\usepackage{empheq}
\renewcommand\theequation{\arabic{equation}}
\usepackage[varg]{txfonts}

\usepackage{epstopdf}

\usepackage{times}

\renewcommand\sout{\bgroup \color{red} \ULdepth=-.5ex \ULset}

\renewcommand{\v}[1]{\textbf{#1}}
\renewcommand{\rm}[1]{\textrm{#1}}
\renewcommand{\d}{\mathrm{d}}

\begin{document}

\title{Investigating Effects of Relativistic Kinematics, Dimensionality, Interactions, and Short-Range Correlations on the Ratio of Quartic over Quadratic Nuclear Symmetry Energies}

\author{Bao-Jun Cai\footnote{bjcai87@gmail.com}}
\affiliation{Quantum Machine Learning Laboratory, Shadow Creator Inc., Shanghai 201208, China}
\author{Bao-An Li\footnote{Bao-An.Li$@$tamuc.edu}}
\affiliation{Department of Physics and Astronomy, Texas A$\&$M
University-Commerce, Commerce, TX 75429-3011, USA}
\date{\today}

\begin{abstract}
While ample evidence for the so-called empirical parabolic law of the Equation of State (EOS) of isospin asymmetric nuclear matter (ANM) has been obtained in many studies within both non-relativistic and relativistic nuclear many-body theories using various interactions, it has been unclear if there is any fundamental physics reason for the small quartic symmetry energy compared to the quadratic one even as the ANM approaches pure neutron matter. 
Within both relativistic and non-relativistic Free Fermi Gas (FFG) models in coordinate spaces of arbitrary dimension $d$ with and without considering Short-Range Correlations (SRC) as well as non-linear Relativistic Mean Field (RMF) models, we study effects of relativistic kinematics, dimensionality, interactions and SRC on the ratio $\Psi(\rho)$ of quartic over quadratic symmetry energies in ANM EOSs.  We found that the ratio $\Psi(\rho)$ in the FFG model depends strongly on the dimension $d$. While it is very small already in the normal 3D space, it could be even smaller in spaces with reduced dimensions for sub-systems of particles in heavy-ion reactions and/or whole neutron stars due to constraints, collectivities and/or symmetries. We also found that the ratio $\Psi(\rho)$ could theoretically become very large only at the ultra-relativistic limit far above the density reachable in neutron stars. On the other hand, 
nuclear interaction directly and/or indirectly through SRC-induced high-momentum nucleons affect significantly the density dependence of $\Psi(\rho)$ compared to the relativistic FFG model prediction. The SRC affects significantly not only the kinetic energy of symmetric nuclear matter but also the ratio $\Psi(\rho)$ while the relativistic corrections are found negligible. Although we found no fundamental physics reason for the $\Psi(\rho)$ to be very small especially at high densities, the results may help better understand the EOS of dense neutron-rich matter. 
\end{abstract}

\maketitle


\def\Ts{\Theta_{\rm{sym}}}
\def\Ps{\Pi_{\rm{sym}}(\chi,\Ts)}
\def\Es{E_{\rm{sym}}}

\section{Introduction and conclusions}\label{S1}

The energy per nucleon $E(\rho,\delta)$ in cold neutron-rich nucleonic matter at density $\rho$ and isospin asymmetry $\delta=(\rho_{\rm{n}}-\rho_{\rm{p}})/\rho$ with $\rho_{\rm{n}}$ and $\rho_{\rm{p}}$ being the neutron and proton densities, respectively, is a basic input for calculating the Equation of State (EOS) for various applications in both nuclear physics and astrophysics. It is usually expanded in even powers of $\delta$ as 
\begin{equation}\label{eos}
E(\rho,\delta)\approx E_0(\rho)+E_{\rm{sym}}(\rho)\delta^2+E_{\rm{sym,4}}(\rho)\delta^4+\mathcal{O}(\delta^{6})
\end{equation}
in terms of the energy per nucleon $E_0(\rho)\equiv E(\rho,0)$ in symmetric nuclear matter (SNM), the symmetry energy (quadratic term) $E_{\rm{sym}}(\rho)\equiv 2^{-1}\partial^2E(\rho,\delta)/\partial\delta^2|_{\delta=0}$, the fourth-order symmetry energy (quatic term) $E_{\rm{sym,4}}(\rho)\equiv {24}^{-1}\partial^4E(\rho,\delta)/\partial\delta^4|_{\delta=0}$, etc.
The first three terms at the nuclear saturation density $\rho_0\approx 0.16\,\rm{fm}^{-3}$, namely $E_0(\rho_0)$, $ E_{\rm{sym}}(\rho_0)$ and $E_{\rm{sym,4}}(\rho_0)$ are empirically constrained to be about $E_0(\rho_0)\approx-16\,\rm{MeV}, E_{\rm{sym}}(\rho_0)\approx 32\,\rm{MeV}$\,\cite{LiBA13,LiBA19,LiBA21} and $E_{\rm{sym,4}}(\rho_0)\lesssim2\,\rm{MeV}$\,\cite{Bom91,Lee98,Ste06,Cai12,Gon17,PuJ17}, indicating that the small-quantity expansion in terms of  $\delta^2$ for the $E(\rho,\delta)$ in asymmetric nuclear matter (ANM) is very effective and sufficient for many physics purposes. 
The small ratio of the quartic over the quadratic symmetry energy terms was confirmed using both the relativistic and non-relativistic phenomenological models as well as several more microscopic many-body theories\,\cite{Lee98,Bom91,Ste06,Cai12,Gon17,PuJ17}. Thus, there are many empirical evidences for the so-called empirical parabolic law of the EOS of neutron-rich matter\,\cite{Bom91}, i.e, safely truncating the expansion of Eq.\,(\ref{eos}) at $\delta^2$. Interestingly, however, certain quantities/processes in neutron stars/astrophysics were found to be significantly affected by even a very small quartic and/or even higher order terms in the expanded $E(\rho,\delta)$, see, e.g., Ref.\,\cite{Cai12} and references therein.

Considering the aforementioned findings and empirical evidences, some interesting questions emerge naturally. For instance, why is the isospin quartic term so small compared with its preceding term? Does there exist any deep physics reason for the seemingly quick convergence in expanding the $E(\rho,\delta)$ while the $\delta$ may be not small (in fact, it may approach 1 not only theoretically but also actually in some regions of neutron stars)? Within most of the energy density functional formalisms in the literature, the kinetic and potential contributions to the $E(\rho,\delta)$ are clearly separated. Generally, the $E(\rho,\delta)$ is written as the sum of nucleon kinetic energies in a Free Fermi Gas (FFG) and their interaction energies evaluated using various nuclear many-body theories. 
The first indication that the (kinetic) quartic symmetry energy may be comparable with the (kinetic) quadratic symmetry energy is from studying effects of the strong isospin dependence of nucleon-nucleon short range correlations (SRCs)\,\cite{Cai16}. The latter leads to a higher fraction of protons compared to neutrons in the high momentum tail (HMT) in the single-nucleon momentum distribution functions in neutron-rich matter as evidenced by findings of several recent SRC experiments\,\cite{Hen14,Due18,Sch19,Sch20}. Incorporating approximately experimental findings regrading the SRC-induced HMTs in a non-relativistic Fermi gas model, the (kinetic) quadratic symmetry energy was found to decrease\,\cite{Xulili,Hen15,LiBA15} while the quartic one increases significantly compared to the FFG model predictions without considering the SRC effects\,\cite{Cai16}. More quantitatively, the kinetic quartic symmetry energy at $\rho_0$ was found to be as large as about 7\,MeV\,\cite{Cai16}. For a recent review, see, e.g., Ref.\,\cite{LCCX18}. It is well known that two-body SRCs in cold nuclear matter push nucleons from below to above the Fermi surface up to about two times the nucleon Fermi momentum. Since the latter is proportional to $\rho^{1/3}$, at high densities relativistic effects naturally become important. Thus, both the SRC and relativistic kinematics are expected to be important for nucleons in the HMT. 
However, it is still unclear how the relativistic kinematics and SRC effects may be intertwined in determining the quartic relative to the quadratic symmetry energy,
i.e.,  $\Psi(\rho)\equiv E_{\rm{sym,4}}(\rho)/E_{\rm{sym}}(\rho)$ (this ratio naturally becomes the ratio of the respective kinetic energies in FFG models). Moreover, nuclear interactions are naturally expected to affect the ratio $\Psi(\rho)$. It is thus also interesting to know the relative kinetic and potential contributions to the various high-order symmetry energies.

In this work, to help shed some new light on the issues mentioned above, we investigate the following topics:
\begin{enumerate}[leftmargin=*,label=(\arabic*)]
\item The ratio $\Psi$ in a non-relativistic FFG model in coordinate spaces of general dimension $d$=1 to 4. We also discuss briefly some possible situations where a sub-system of particles during heavy-ion reactions or all particles in whole neutron stars can be considered approximately as moving in 1D or 2D due to constraints, collectivities and/or symmetries.
\item The ratio $\Psi$ in a 3D ($d$=3) relativistic FFG model.
\item Nuclear interaction effects on the ratio $\Psi$ within a nonlinear relativistic mean field (RMF) model with respect to the relativistic FFG model predictions.
\item Relativistic corrections to the first four terms in the kinetic EOS of neutron-rich nucleonic matter and the ratio $\Psi$ in the presence of SRC-induced HMT in the single-nucleon momentum distribution. 
\end{enumerate}
Our main findings can be summarized as follows:
\begin{enumerate}[leftmargin=*,label=(\arabic*)]
\item The ratio $\Psi$ in the FFG model depends strongly on the dimension $d$ of the coordinate space in which nucleons move around. While in the normal 3D space, the ratio $\Psi$ is very small, it can be even smaller or much larger in spaces with reduced or extended dimensions. The smallness of $\Psi$ in the conventional 3D space is thus nothing special. Moreover, relativistic corrections to the kinetic energy of SNM as well as the quadratic and quartic symmetry energies all depend strongly on the dimensionality of the system considered. 
\item The ratio $\Psi$ could become very large if either the density is large enough (in the ultra-relativistic limit) or nucleon-nucleon interactions are taken into account (either indirectly via the SRC-induced HMT or directly through the effective nuclear interactions). 
\item In the presence of the SRC-induced HMT, relativistic corrections to both the quadratic and quartic kinetic symmetry energies are small mainly due to the small $k_{\rm{F}}/M$ ratio with $k_{\rm{F}}$ and $M$ being the Fermi momentum and nucleon mass, respectively.
\end{enumerate}

\section{$E_{\textmd{sym},4}^{\textmd{kin}}(\rho)/E^{\textmd{kin}}_{\textmd{sym}}(\rho)$ in Free Fermi Gas models}\label{S2}

\subsection{Non-relativistic FFG Model predictions in 3D}\label{SEC_II_A}
As a useful reminder of the basic formalisms and notations used in the present work, we briefly recall here predictions of the non-relativistic FFG model in a 3D coordinate space. 
For nucleons in ANM, the FFG single-nucleon momentum distributions are step functions, i.e., $n_{\v{k}}^J(\rho,\delta)=\Theta(k_{\rm{F}}^J-|\v{k}|)$ where $J=\rm{n}$ and $\rm{p}$.  
The nucleon density $\rho_{J}$ and the corresponding Fermi momentum $k_{\rm{F}}^J$ are related by $
k_{\rm{F}}^J=k_{\rm{F}}(1+\tau_3^J\delta)^{1/3}$,
where $\tau_3^{\rm{n}}=+1$ and $\tau_3^{\rm{p}}=-1$ are the third components of the isospin vector of neutrons and protons, respectively.
The average kinetic energy per nucleon in ANM is then straightforwardly obtained as
\begin{align}\label{def_EOSANMFFG}
E^{\rm{kin}}(\rho,\delta)=&\left.\left[\int_0^{k_{\rm{F}}^{\rm{n}}}\frac{\v{k}^2}{2M}\d\v{k}+\int_0^{k_{\rm{F}}^{\rm{p}}}\frac{\v{k}^2}{2M}\d\v{k}\right]\right/2\int_0^{k_{\rm{F}}}\d\v{k}\notag\\
=&\frac{3}{5}\frac{k_{\rm{F}}^2}{2M}\frac{1}{2}\left[(1+\delta)^{5/3}+(1-\delta)^{5/3}\right]\notag\\
\approx&\frac{3k_{\rm{F}}^2}{10M}+\frac{k_{\rm{F}}^2}{6M}\delta^2+\frac{k_{\rm{F}}^2}{162M}\delta^4
+\frac{7k_{\rm{F}}^2}{4374M}\delta^6.
\end{align}
One can then find that $E_0^{\rm{kin}}(\rho)=3k_{\rm{F}}^2/10M$, $E_{\rm{sym}}^{\rm{kin}}(\rho)
={k_{\rm{F}}^2}/{6M}$, $E_{\rm{sym,4}}^{\rm{kin}}(\rho)={k_{\rm{F}}^2}/{162M}$, and $E_{\rm{sym,6}}^{\rm{kin}}(\rho)=7k_{\rm{F}}^2/4374M$.
The ratio $\Psi\equiv E^{\rm{kin}}_{\rm{sym,4}}(\rho)/E^{\rm{kin}}_{\rm{sym}}(\rho)$ is 1/27. 
By using the Fermi momentum $k_{\rm{F}}=(3\pi^2\rho/2)^{1/3}\approx263\,\rm{MeV}$ corresponding to the saturation density $\rho_0\approx0.16\,\rm{fm}^{-3}$, we obtain $E_0^{\rm{kin}}(\rho_0)\approx 22.2\,\rm{MeV}$, $E_{\rm{sym}}^{\rm{kin}}(\rho_0)\approx 12.3\,\rm{MeV}$, $ E_{\rm{sym,4}}^{\rm{kin}}(\rho_0)\approx 0.45\,\rm{MeV}$, and $E_{\rm{sym},6}^{\rm{kin}}(\rho)\approx 0.12\,\rm{MeV}$, respectively.

The above well known predictions by the non-relativistic FFG model serve as useful references in evaluating effects of the relativistic kinematics and SRC as well as nuclear interactions missing in this model. For example, 
since $E_0^{\rm{kin}}(\rho_0)\approx 22.2\,\rm{MeV}$, to reproduce the empirical binding energy of about $-16\,\rm{MeV}$ at $\rho_0$, an interaction contribution of about $-38.2$\,MeV is required. Similarly, since the kinetic symmetry energy is about 12.3\,MeV, an interaction contribution of about 20\,MeV would be needed to reproduce the total empirical nuclear symmetry energy of about 32\,MeV at $\rho_0$\,\cite{LiBA13}.
While this model predicts a fourth-order kinetic symmetry energy of about 0.45\,MeV and several microscopic theories and phenomenological models predict 
a total $E_{\rm{sym,4}}(\rho_0)$ less than about 2\,MeV\,\cite{Lee98, Bom91,Ste06,Cai12,Gon17,PuJ17}, there is currently no community census on the empirical value of $E_{\rm{sym,4}}(\rho_0)$. 
Needless to say, the situation is worse for the sixth-order symmetry $E_{\rm{sym},6}^{\rm{kin}}(\rho_0)$ (both the kinetic term and the total). We thus can not judge whether the relative kinetic contribution to the total symmetry energy is decreasing or not going from the $\delta^2$ to higher-order terms.

\subsection{FFG Model predictions for $\langle k^{\sigma}\rangle$ with $\sigma\ge 2$ and $E_{\textmd{sym},4}^{\textmd{kin}}(\rho)/E^{\textmd{kin}}_{\textmd{sym}}(\rho)$ in $d$-Dimensions}\label{d-space}
As shown above, the FFG model predicts that the ratio $E^{\rm{kin}}_{\rm{sym,4}}(\rho)/E^{\rm{kin}}_{\rm{sym}}(\rho)=1/27$. To put this result in appropriate perspective,  noticing that 
the relativistic kinetic energy $T(\v{k})=\sqrt{\v{k}^2+M^2}-M$ of a nucleon can be expanded as
\begin{align}\label{def_nonexp}
T(\v{k})\approx\frac{\v{k}^2}{2M}\left(1-\frac{\v{k}^2}{4M^2}+\frac{\v{k}^4}{8M^4}-\frac{5\v{k}^6}{64M^6}+\frac{7\v{k}^8}{128M^8}\right),
\end{align}
to order $\v{k}^{10}$, here we first consider mathematically ANM in a coordinate space of arbitrary dimension $d$ and calculate the average $\langle k^{\sigma}\rangle$ of $k^{\sigma}$ with $\sigma\ge 2$ being an integer.
The $\langle k^{\sigma}\rangle$ is a function of density and isospin asymmetry,  i.e., 
\begin{align}\label{k-sigma}
\langle k^{\sigma}(\rho,\delta)\rangle\equiv&\left.\left[\int_0^{k_{\rm{F}}^{\rm{n}}}k^{\sigma}\d^d\v{k}+\int_0^{k_{\rm{F}}^{\rm{p}}}k^{\sigma}\d^d\v{k}\right]\right/2\int_0^{k_{\rm{F}}}\d^d\v{k}\notag\\
=&\Upsilon\left[(1+\delta)^{1+\sigma/d}+(1-\delta)^{1+\sigma/d}\right],
\end{align}
where $\sigma\in \rm{Z}$ and $\Upsilon$ a coefficient independent of $\delta$. We emphasize that the power $\sigma$ and dimension $d$ appear in the combination of $\sigma/d$.
We also notice that the case $\sigma=2$ corresponds to the conventional non-relativistic kinetic energy, and $\sigma=4$ (with the energy scales as $k^4/M^3$) induces the first relativistic correction and $\sigma=6$ induces the second relativistic correction, etc, as indicated by the expansion in Eq.\,(\ref{def_nonexp}).
In addition, $k_{\rm{F}}^J=k_{\rm{F}}(1+\tau_3^J\delta)^{1/d}$ is the nucleon Fermi momentum with the Fermi momentum in symmetric matter given by $k_{\rm{F}}(\rho)=[\rho2^{d-2}\pi^{d/2}\Gamma(d/2+1)]^{1/d}\sim\rho^{1/d}$.

The quantity $\langle k^{\sigma}(\rho,\delta)\rangle$ can be expanded around $\delta=0$ as
\begin{eqnarray}
\langle k^{\sigma}(\rho,\delta)\rangle&=&\sum_{j=0}k_{\rm{sym},2j}^{\sigma}(\rho)\delta^{2j}\notag\\
&\approx& k_0^{\sigma}(\rho)+k_{\rm{sym}}^{\sigma}(\rho)\delta^2+k^{\sigma}_{\rm{sym,4}}(\rho)\delta^4
+\mathcal{O}(\delta^{6})
\cdots, 
\end{eqnarray}
where $k_0^{\sigma}(\rho)\equiv k_{\rm{sym},0}^{\sigma}(\rho)$ and $k_{\rm{sym}}^{\sigma}(\rho)\equiv k_{\rm{sym,2}}^{\sigma}(\rho)$. Then from Eq.\,(\ref{k-sigma}), it is straightforward to obtain the ratio 
\begin{align}\label{def_pp1}
\Psi^{\sigma}_{d}(j=2)\equiv\left.{k^{\sigma}_{\rm{sym,4}}(\rho)}\right/{k^{\sigma}_{\rm{sym}}(\rho)}
=\frac{1}{12}\left(\frac{\sigma}{d}-2\right)\left(\frac{\sigma}{d}-1\right).
\end{align}
Depending on the value of $\sigma/d$, this ratio is not necessarily as small as the $E^{\rm{kin}}_{\rm{sym,4}}(\rho)/E^{\rm{kin}}_{\rm{sym}}(\rho)=1/27$ in the FFG model. For example, if one takes $\sigma=4$ and $d=1$,  i.e., the average of the momentum to the fourth power (corresponding to the first relativistic correction) in dimension 1,  one obtains $\Psi^4_1(j=2)=1/2$.
Moreover, from the above relation, one has
\begin{equation}\label{k42}
\lim_{d\to\infty}\frac{k^{\sigma}_{\rm{sym,4}}(\rho)}{k^{\sigma}_{\rm{sym}}(\rho)}=\frac{1}{6},
\end{equation}
i.e., in an imagined space of infinite dimensions the ratio $\Psi$ takes the value of $1/6\approx16.7$\%, which is independent of the power $\sigma$ and is about five times the FFG model prediction in 3D.
As $d\to\infty$ (while fixing $\sigma$),  the isospin dependence of $\langle k^{\sigma}(\rho,\delta)\rangle$ is weak (approaching zero) as shown in Eq. (\ref{k-sigma}). However, these isospin-expansion terms are not strictly zero. In particular, we have $k_{\rm{sym,4}}^{\sigma}(\rho)=\Upsilon(g/6-g^2/12-g^3/6+g^4/12)$ and $k_{\rm{sym}}^{\sigma}(\rho)=\Upsilon(g+g^2)$ and both approach zero as $g=\sigma/d\to0$ in the infinite-$d$ limit, leading to a finite ratio of $1/6$ between $k_{\rm{sym,4}}^{\sigma}(\rho)$ and $k_{\rm{sym,4}}^{\sigma}(\rho)$ as demonstrated in Eq. (\ref{k42}). Thus, the special relation (\ref{k42}) and the general expression (\ref{k-sigma}) are consistent with each other.

 \renewcommand*\figurename{\small Fig.}
\begin{figure}[h!]
\centering
\includegraphics[width=2.cm]{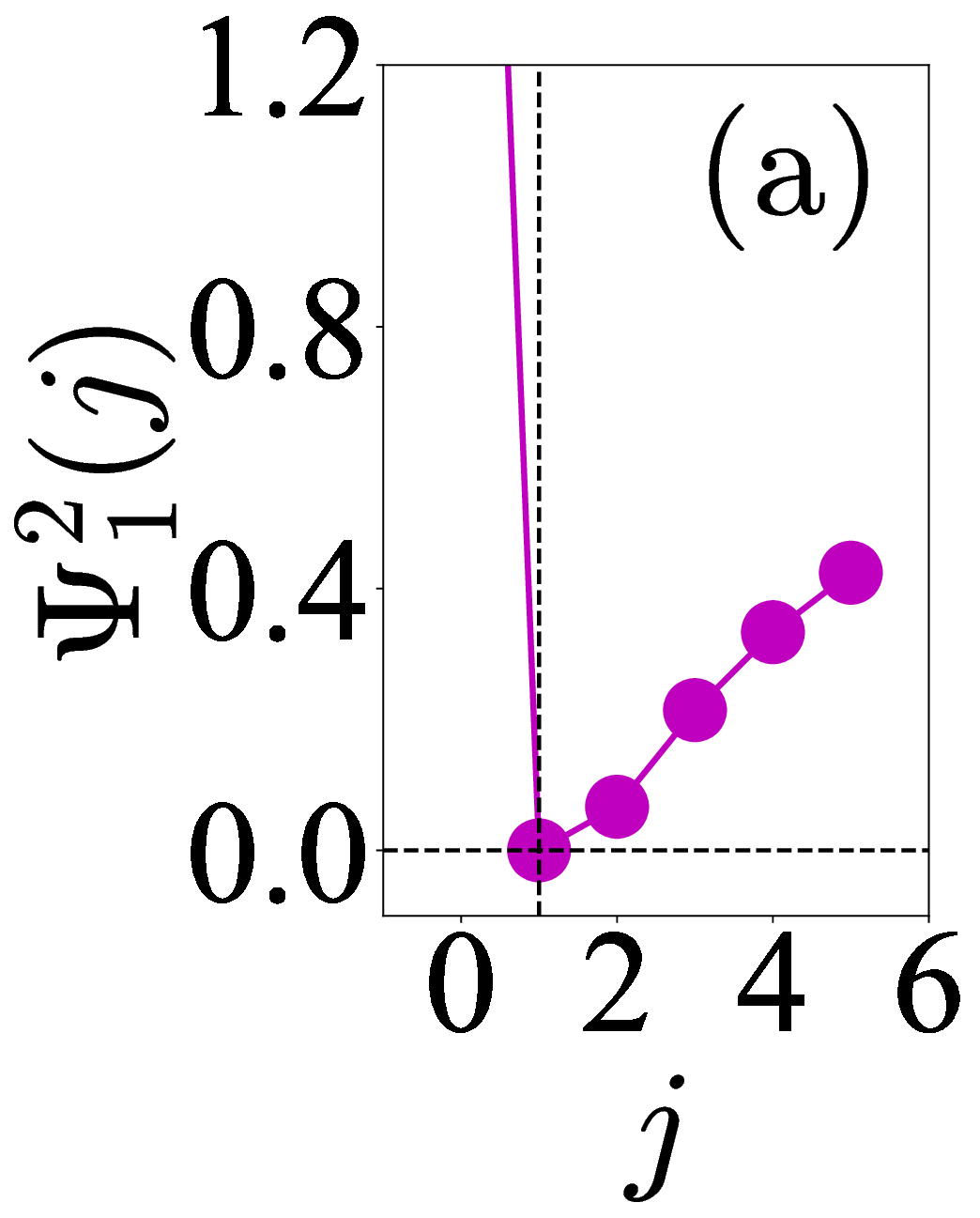}
\includegraphics[width=2.cm]{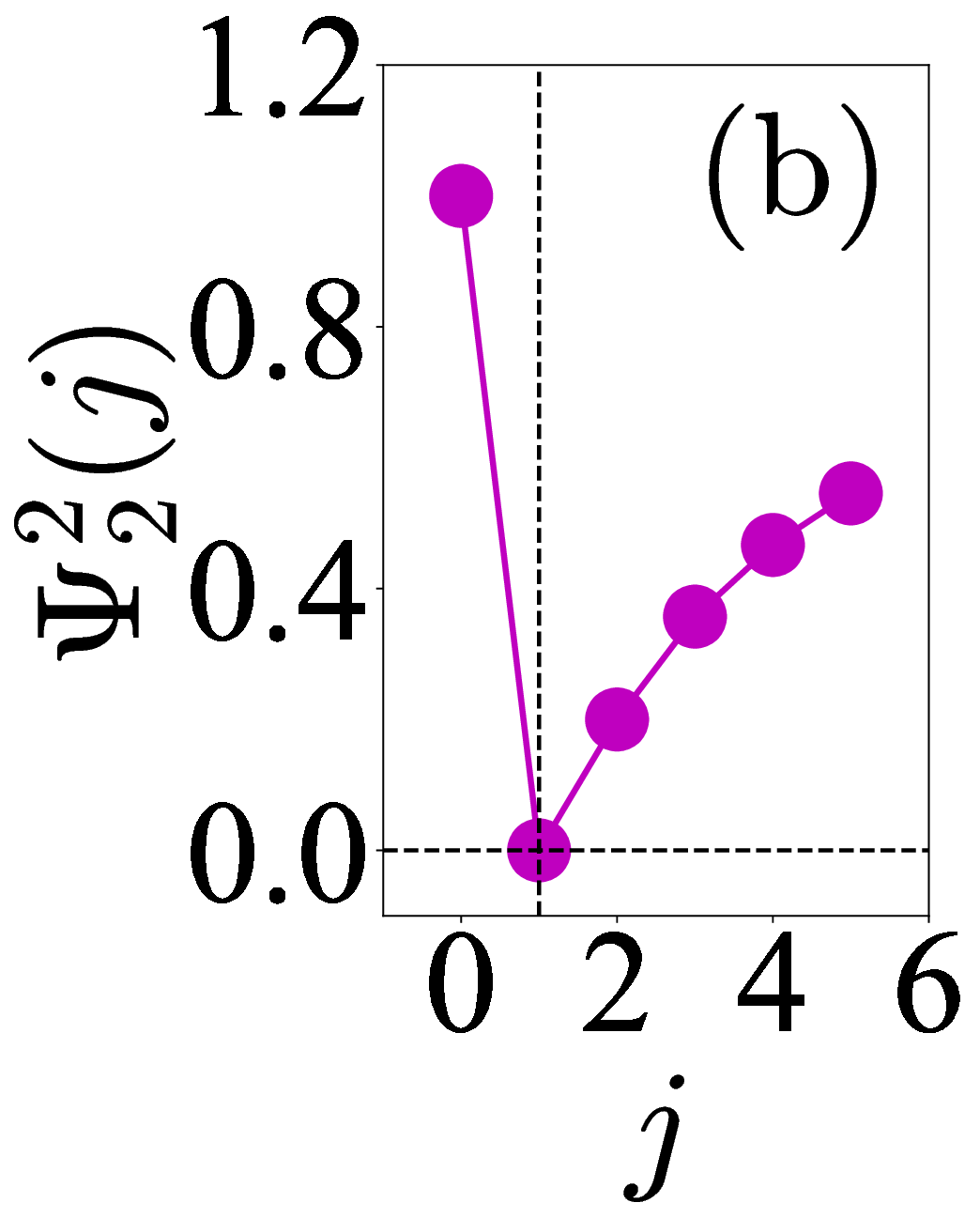}
\includegraphics[width=2.cm]{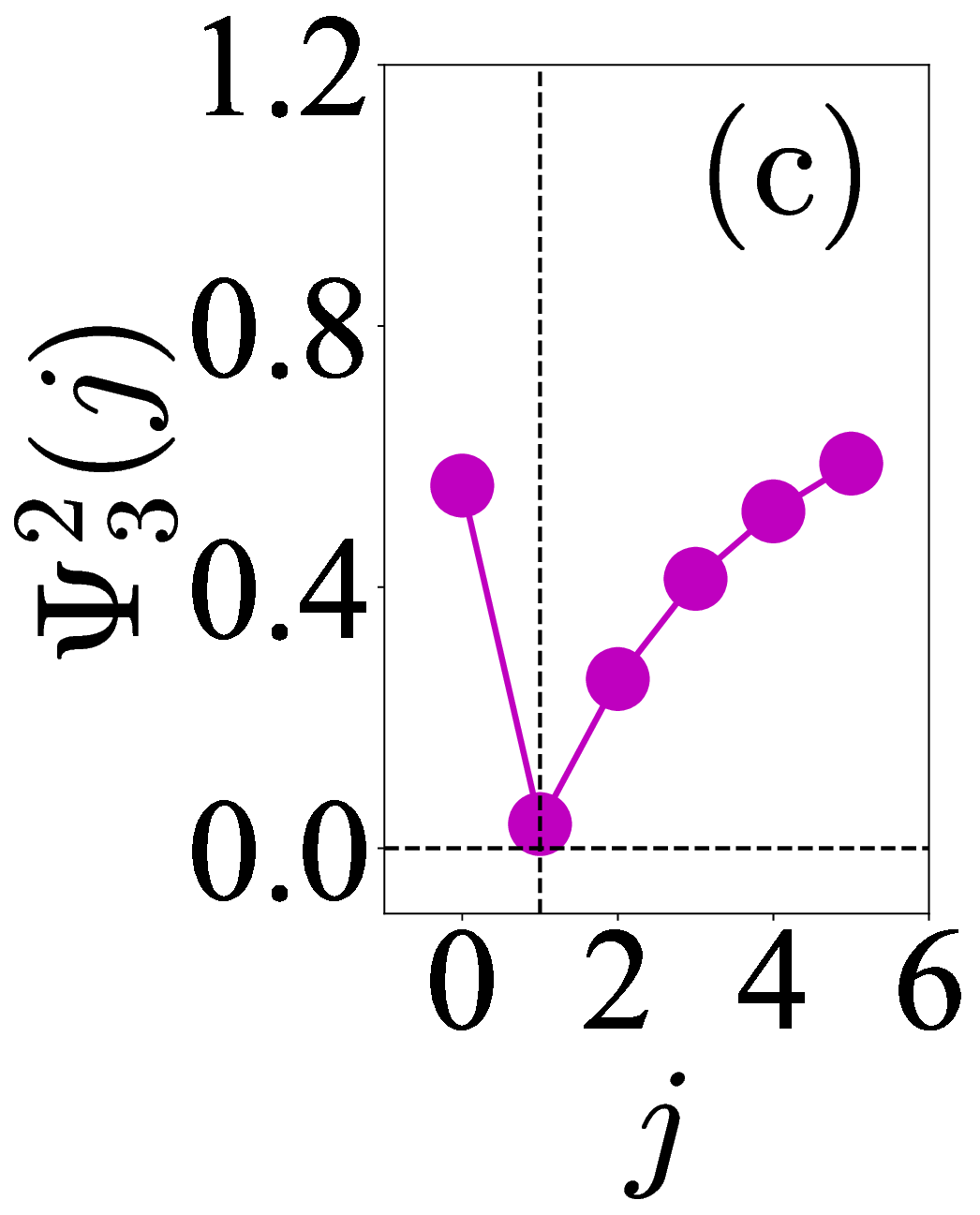}
\includegraphics[width=2.cm]{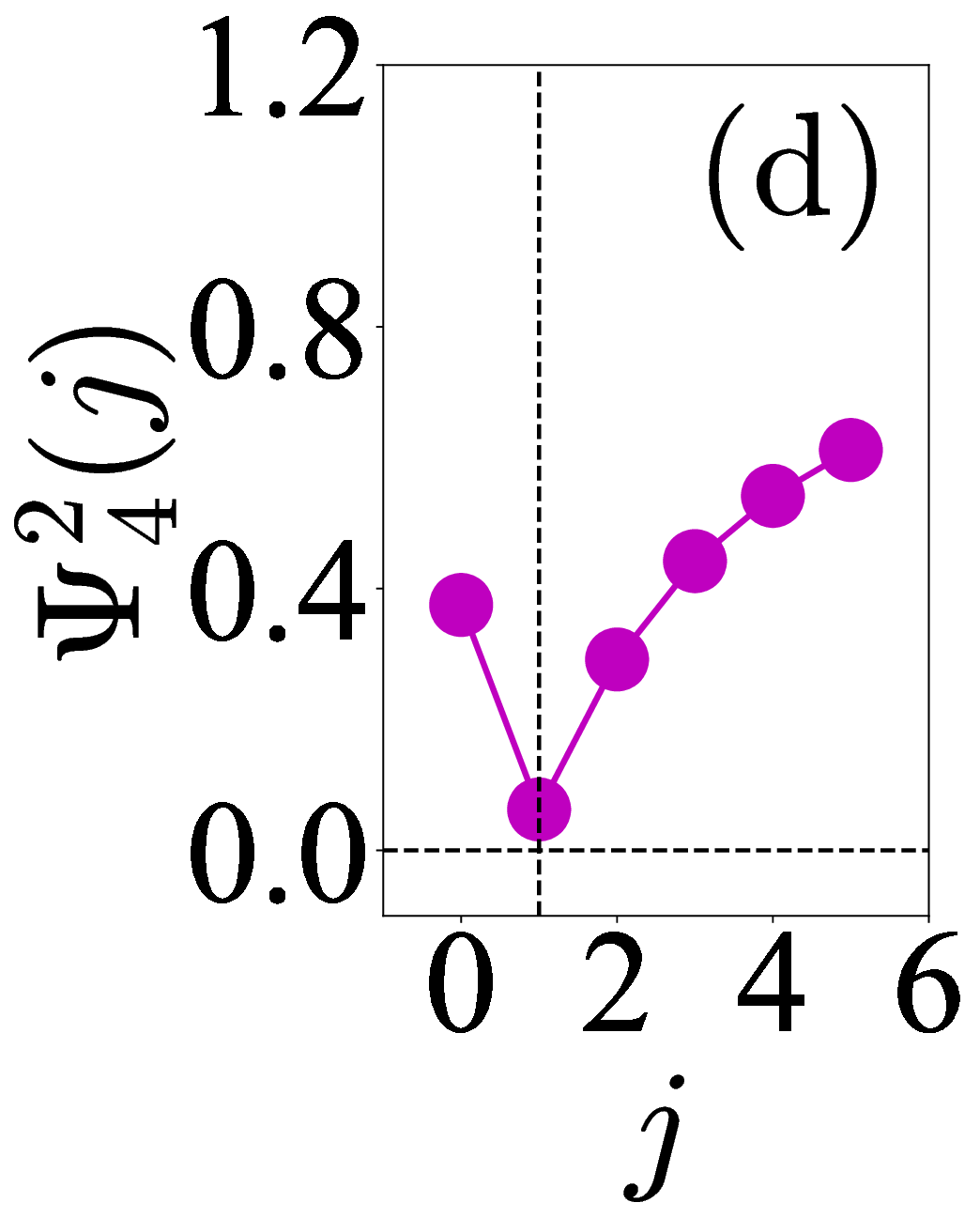}\\
\includegraphics[width=2.cm]{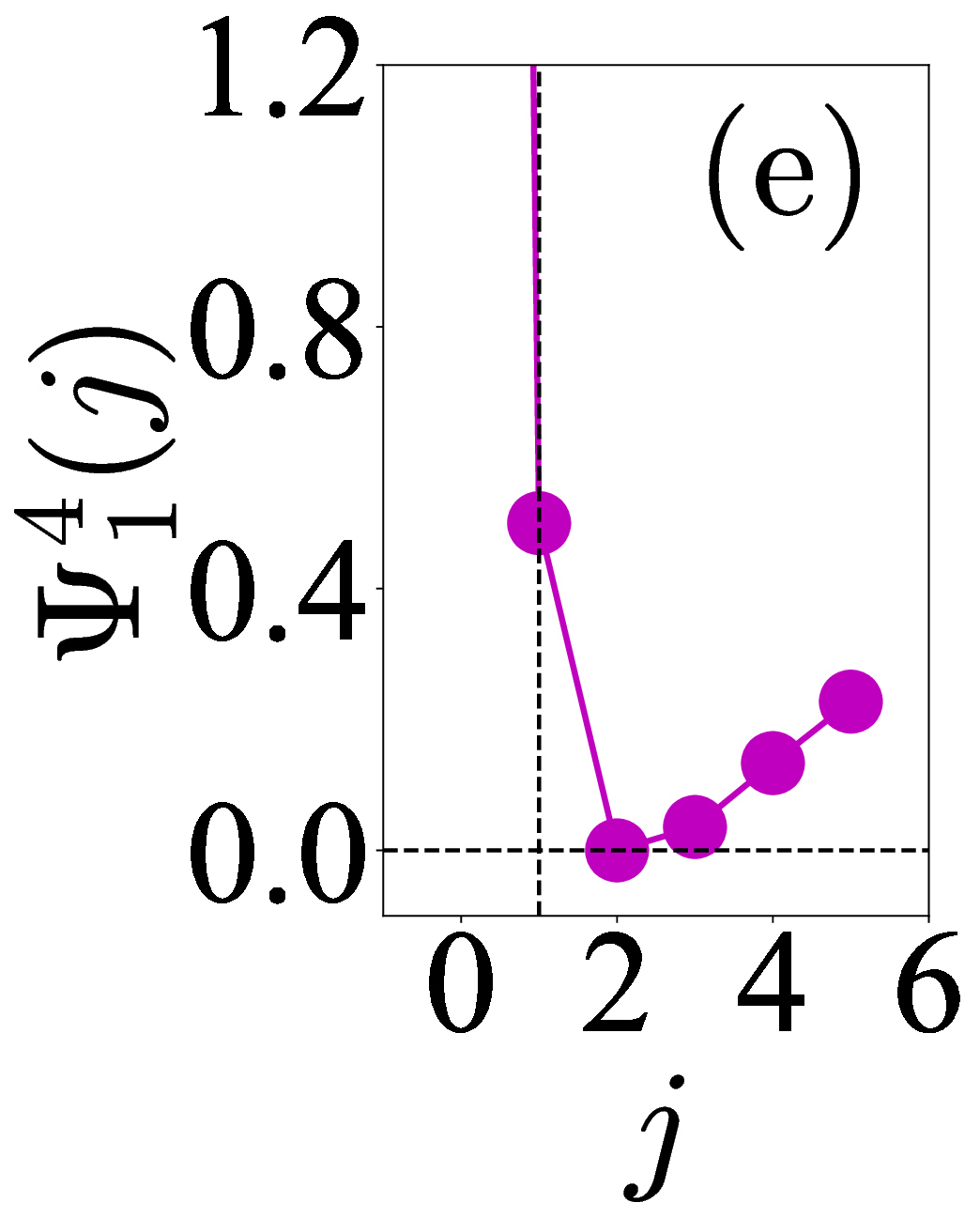}
\includegraphics[width=2.cm]{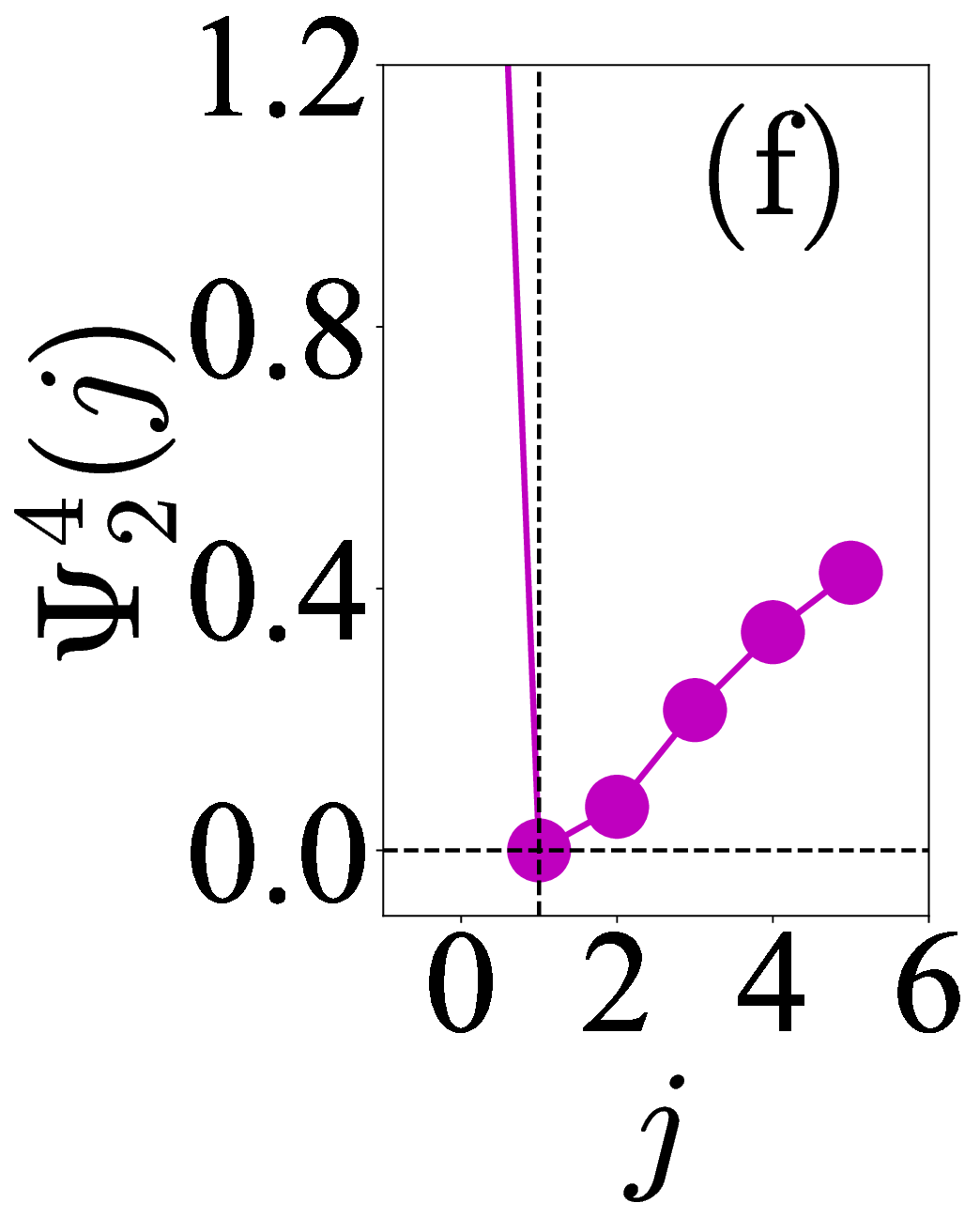}
\includegraphics[width=2.cm]{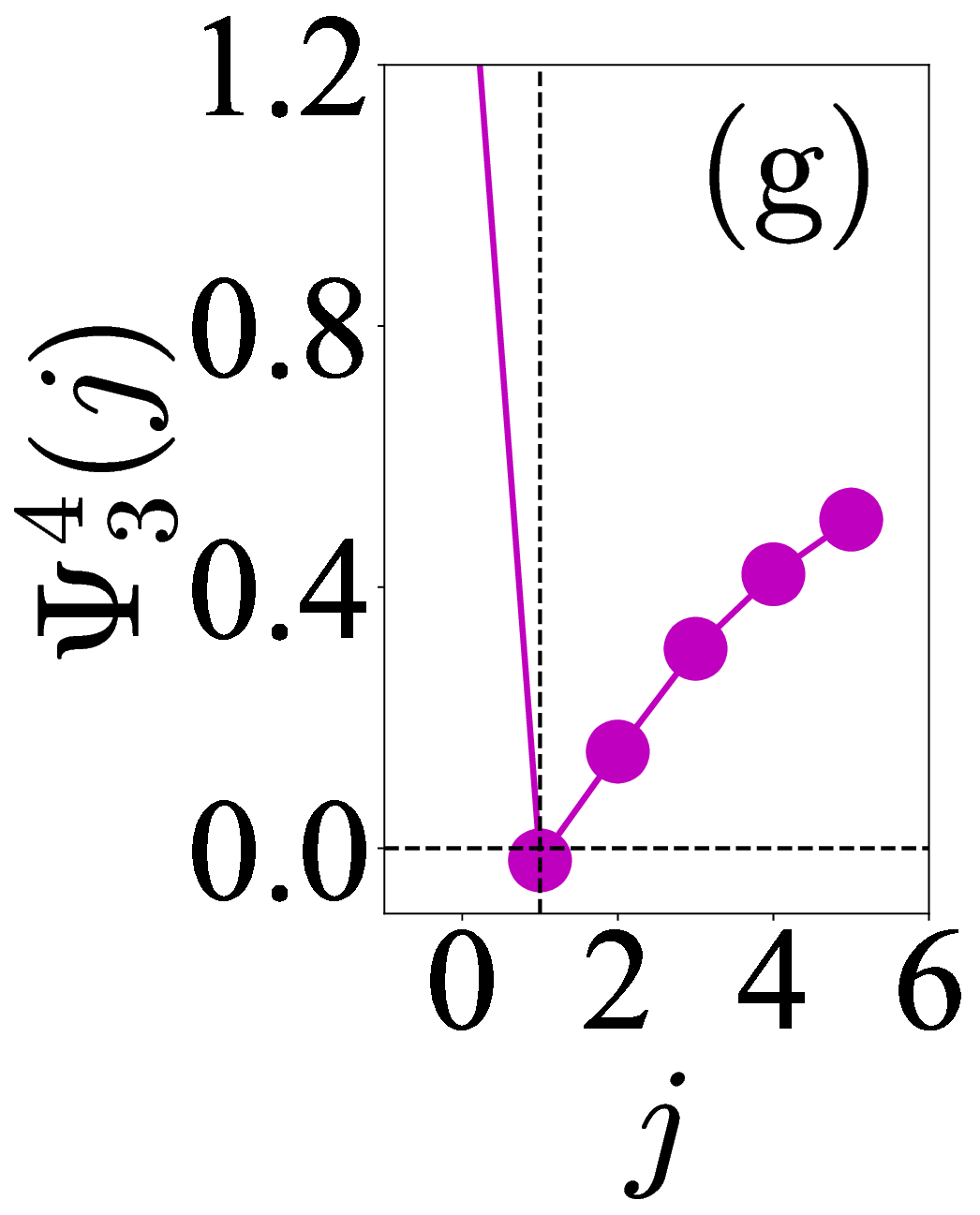}
\includegraphics[width=2.cm]{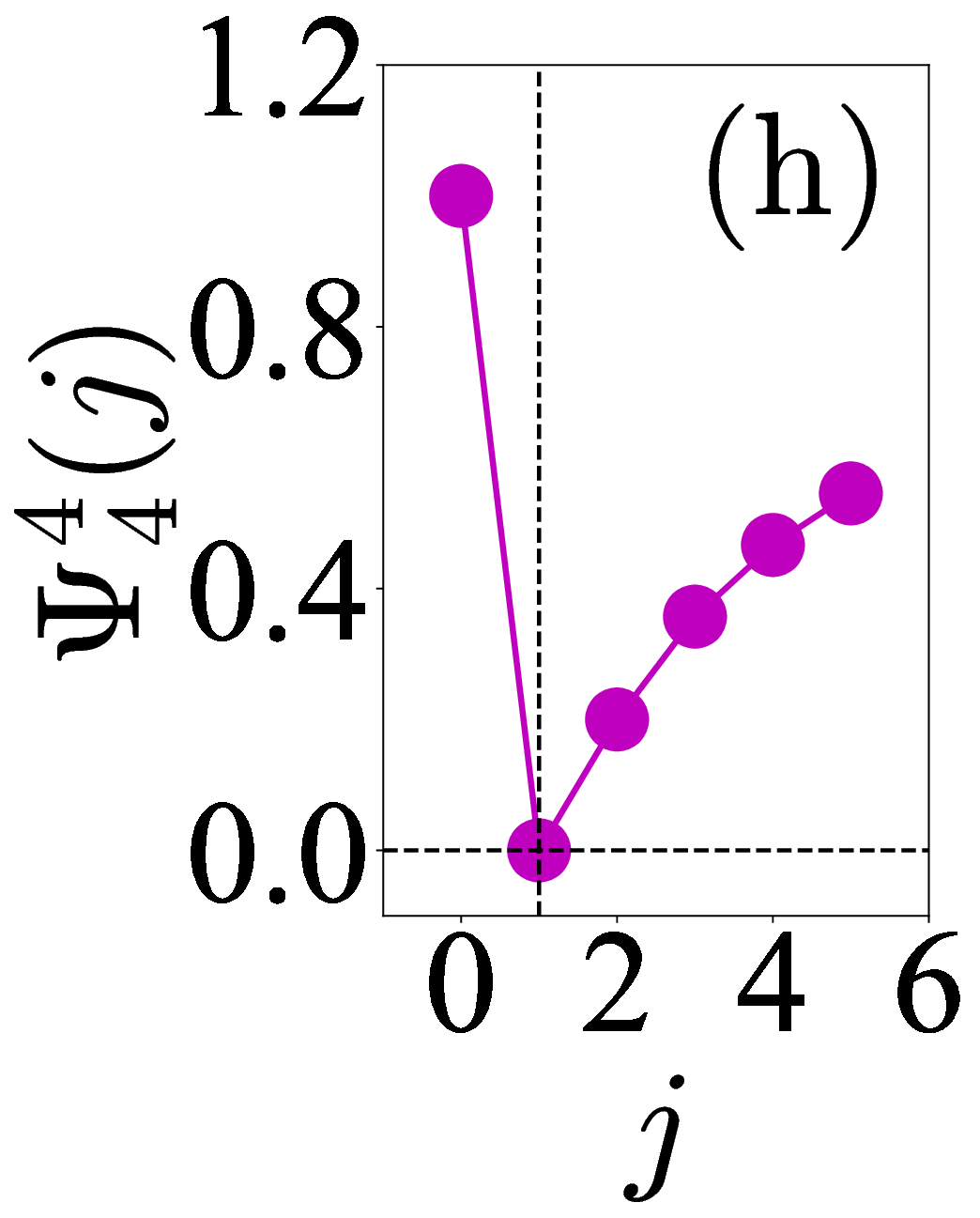}\\
\includegraphics[width=2.cm]{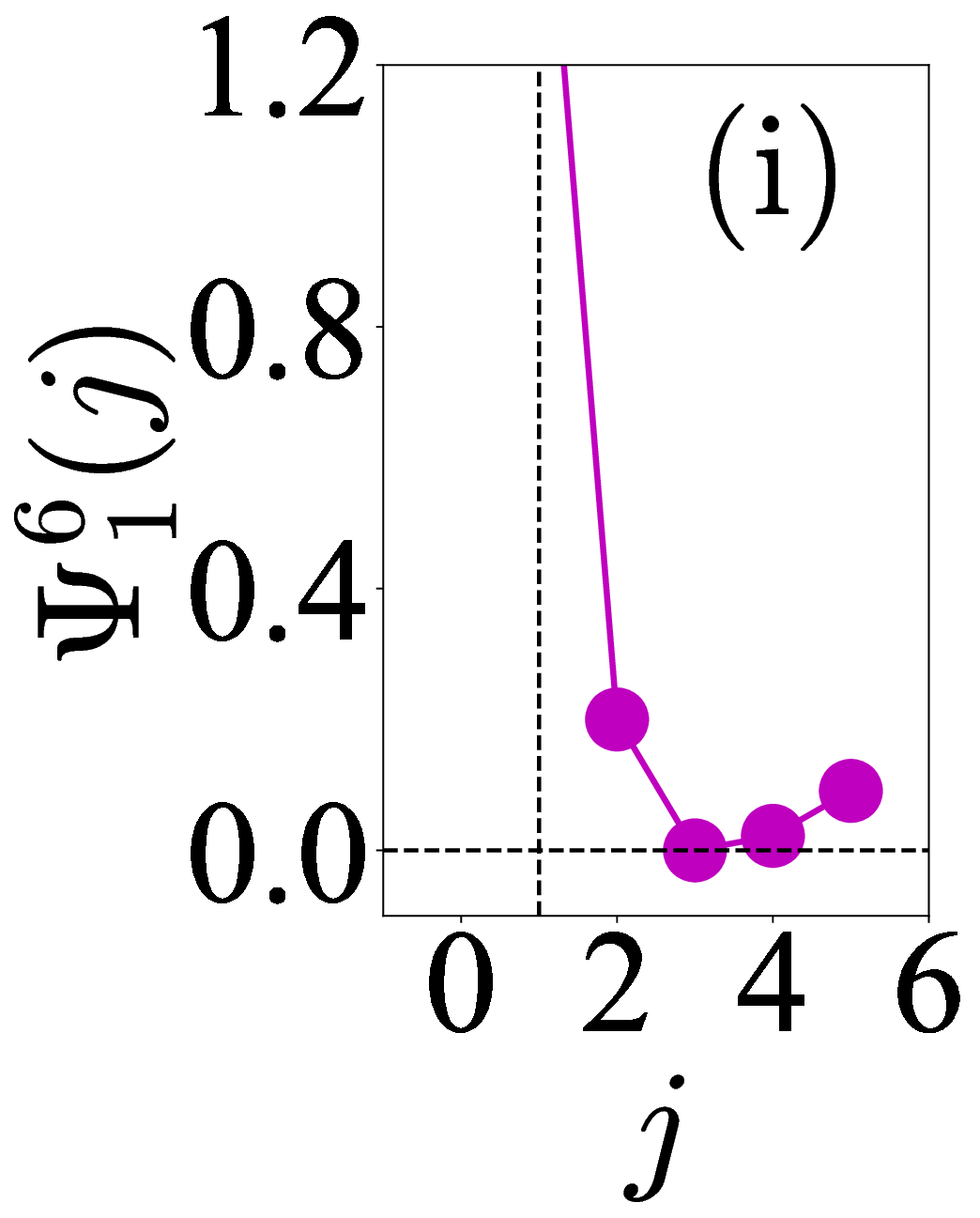}
\includegraphics[width=2.cm]{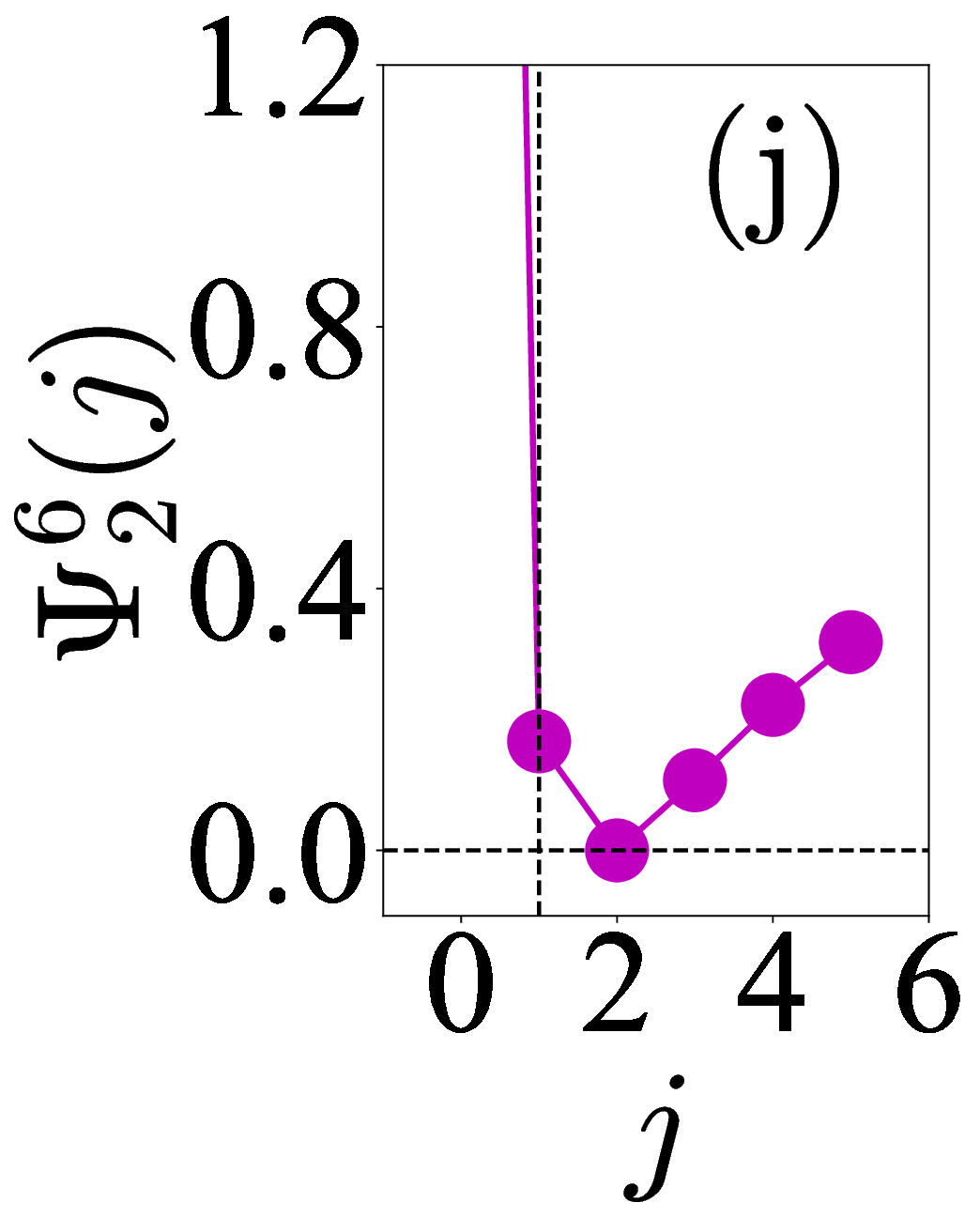}
\includegraphics[width=2.cm]{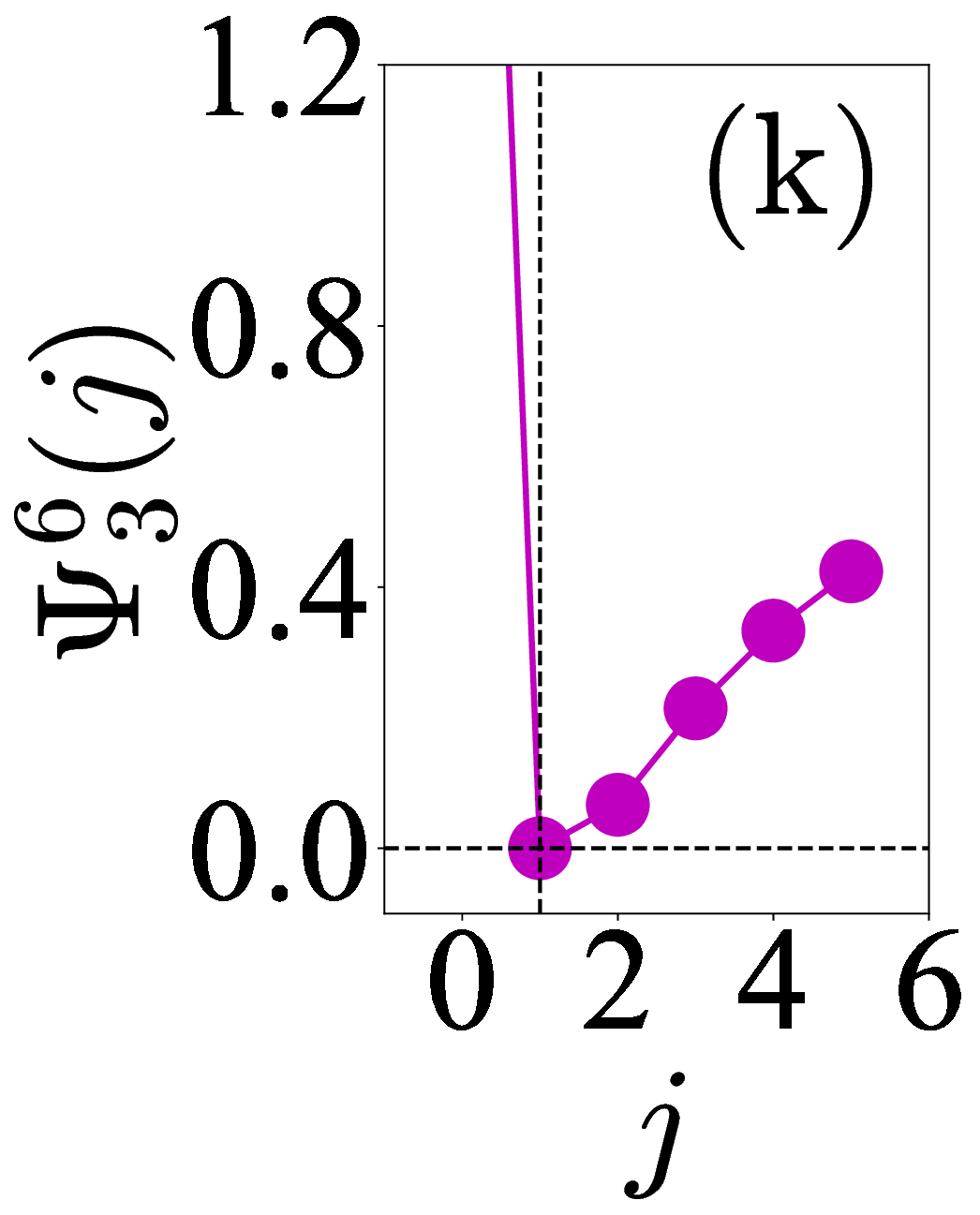}
\includegraphics[width=2.cm]{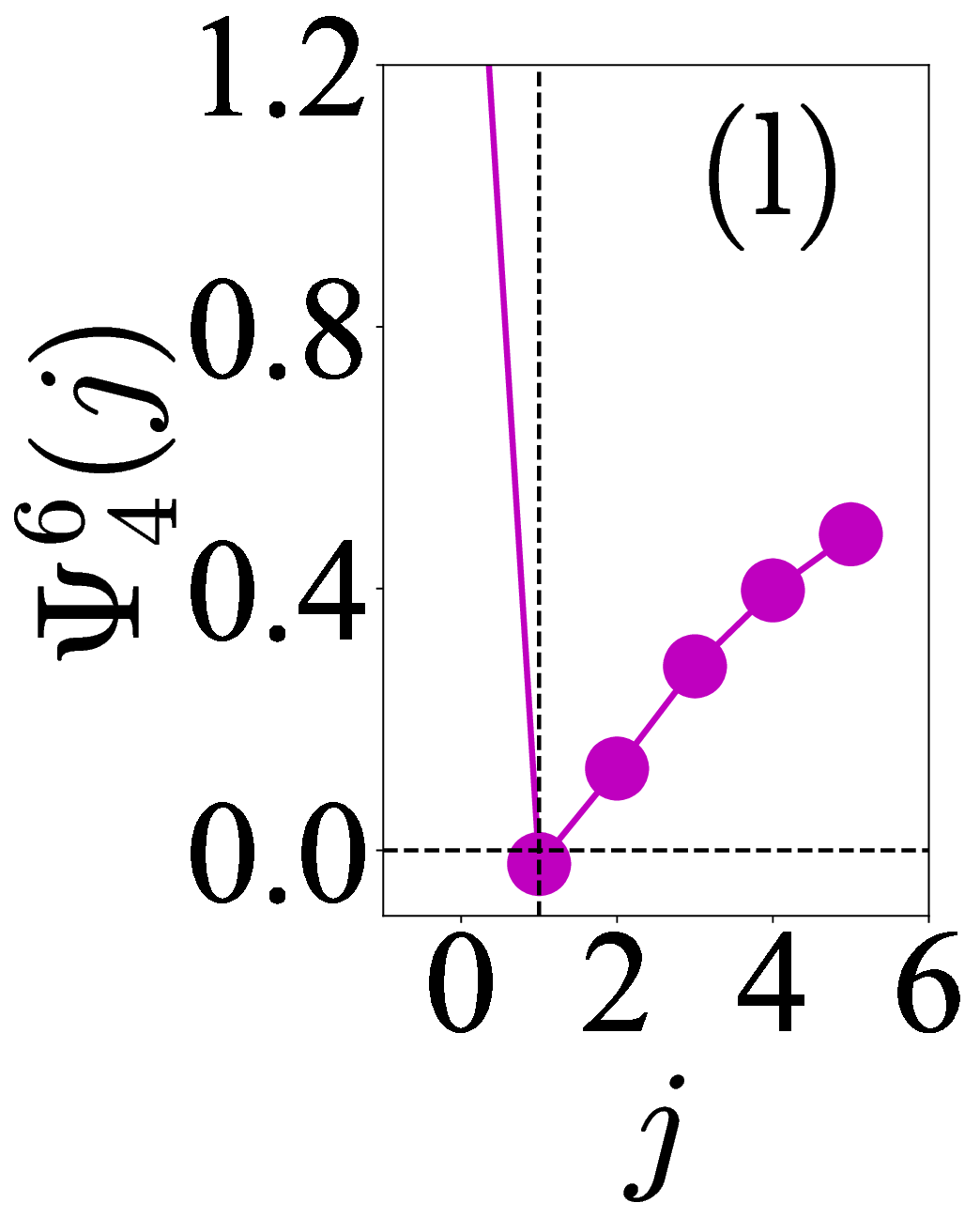}
 \caption{(Color Online). The ratio $\Psi_d^{\sigma}(j)=k^{\sigma}_{\rm{sym},2j+2}/k^{\sigma}_{\rm{sym},2j}$ as a function of $j$ with different $d$ by fixing $\sigma=2$ (first line from (a) to (d)), $\sigma=4$ (second line from (e) to (h)) and $\sigma=6$ (third line from (i) to (l)).
The vertical dashed black line corresponds to $j=1$.}\label{fig_RatioPsi}
\end{figure}

More generally,  one can obtain the ratio between any two adjacent terms in expanding the $\langle k^{\sigma}(\rho,\delta)\rangle$ as a function of $\delta^{2j}$. Specifically, the ratio of the coefficient of the 
$(2j+2)$-term and the $2j$-term, i.e., ${k^{\sigma}_{\rm{sym},2j+2}}/{k^{\sigma}_{\rm{sym},2j}}\equiv\Psi_d^{\sigma}(j)$ (thus the $\Psi_d^{\sigma}(j=2)$ defined in (\ref{def_pp1}) is a special case) is
\begin{align}\label{def_Psi}
\Psi_d^{\sigma}(j)=\frac{1}{(2j+1)(2j+2)}\left(\frac{\sigma}{d}-2j+1\right)\left(\frac{\sigma}{d}-2j\right).
\end{align}
In Fig.\,\ref{fig_RatioPsi} the ratio $\Psi_d^{\sigma}(j)$ as a function of $j$ with different $d$ (from 1 to 4) is shown by fixing $\sigma=2$ (first row from (a) to (d), conventional kinetic energy), $\sigma=4$ (second row from (e) to (h), first relativistic correction) and $\sigma=6$ (third line from (i) to (l), second relativistic correction). From the first row of Fig.\,\ref{fig_RatioPsi} (panels (a) to (d)),  one can see that the ratio of the kinetic quartic over quadratic symmetry energies (characterized by $j=1$) is a very ``special'' point (at the bottom of the curves) irrespective of the dimension $d$.
In particular, for the special case of $\sigma=2$ and $d=3$ (kinetic energy or the square of the momentum in dimensions 3), the ratio $\Psi_3^2(j)$ takes the form $
{k^{\sigma}_{\rm{sym},2j+2}}/{k^{\sigma}_{\rm{sym},2j}}=({18j^2-21j+5})/({18j^2+27j+9})$, 
while if $\sigma=d=2$ (i.e., the square of the momentum in dimensions 2), then ${k^{\sigma}_{\rm{sym},2j+2}}/{k^{\sigma}_{\rm{sym},2j}}=(2j^2-3j+1)/(2j^2+3j+1)$.

Similarly, the corresponding ratio from the first relativistic correction to the kinetic EOS (the average of $k^4$) is shown in the second row of Fig.\,\ref{fig_RatioPsi} with different $d$ by fixing $\sigma$ at four.
Except the case $(\sigma,d)=(4,1)$ (panel (e)),  all the other three sets (panels (f) to (h)) predict that the minimum of $\Psi_d^4$ occurs at $j=1$. 
The second relativistic correction effect is shown in the third row of Fig.\,\ref{fig_RatioPsi}, here for $d\leq 2$ the minimum of $\Psi_d^6$ occurs at $j\neq 1$.
From these graphs, one finds that $j=1$ is in fact a special point with respect to the variation of the dimension $d$ (around 3) and power $\sigma$ (around 2). Since the $\sigma$ and dimension $d$ appear in the combination of $\sigma/d$ in $\Psi_d^{\sigma}(j)$, the panels (a), (f) and (k) of Fig.\,\ref{fig_RatioPsi} are identical.

Generally, we have from Eq.\,(\ref{def_Psi}) that
\begin{equation}
\lim_{j\to\infty}{k^{\sigma}_{\rm{sym},2j+2}}/{k^{\sigma}_{\rm{sym},2j}}=1.
\end{equation} 
It indicates that the radius of convergence of the expansion of the kinetic EOS in terms of $\delta$ is simply $R_{\delta}=1$\,\cite{Wel16}, irrespective of the integer $\sigma$.
For $\sigma=0$, the expression (\ref{def_Psi}) gives a nonzero value, i.e.,  $\lim_{\sigma\to0}\Psi_d^{\sigma}(j)=(2j-1)(2j)/(2j+1)(2j+2)$, although in this case both the $k_{\rm{sym},2j+2}^{\sigma}$ and $k_{\rm{sym},2j}^{\sigma}$ are approaching zero (but their ratio is not necessarily zero).

To this end, it is worth noting that all up-to-today calculations of ANM EOS are performed in dimensions 3. Then, why is it physically meaningful and possibly useful practically besides mathematically doable as we have shown above to consider the kinetic EOS in spaces with dimensions other than 3? First of all, we notice that constraints, symmetries and/or collectivities can all reduce the minimum number of degrees of freedom or the dimensions of the coordinate space necessary to fully describe a many-particle system. In fact, many exciting novel features/breakthroughs have been established in systems with reduced dimensions, see, e.g., Refs.\,\cite{And82,Sar11,Pit16} for reviews. New experimental techniques have been developed to generate not only pure 1D or 2D systems but also two-component many-particle systems with mixed dimensions.  
For instance, the dimensionality of space for cold atoms can be changed by means of strong optical lattices\,\cite{Blo08,Lew07,Lew12,Gro17}. Interesting new physics has been revealed from studying 1D and 2D systems\,\cite{dD1,dD2,dD3,dD4,Pet16,Pet00,Mor15,Ber11}, such as, a two-species Fermi gas with mixed dimensions (one component in 1D or 2D while the other one in 3D)\,\cite{dD5}, a three-dimensional resonant Bose-Fermi mixture at zero temperature\,\cite{Ber13} as well as the two-dimensional Fermi-Bose dimers with a stable $p$-wave resonant interaction\,\cite{Baz18}.  Moreover, new physics insights regarding the relationship between the two-body energy spectrum and the scattering phase shifts can be obtained by studying two particle scatterings in a $d$-dimensional space\,\cite{Tan}. It was also shown that a multiple-body problem could be mapped to a two-body problem in a higher dimension\,\cite{Guo}. Thus, explorations of many-body systems in spaces with different dimensions may provide important insights into some physics problems. 

While we are not aware of any study about the EOS in spaces other than 3D in nuclear physics and/or astrophysics, we notice that the production of particle jets or their squeeze-out from the participant region in a specific direction may effectively create an approximately 1D sub-system while the collective flow in the reaction plane may create approximately a 2D sub-system in intermediate-relativistic energy heavy-ion reactions. 
Moreover, neutron star mergers may create bursts of $\gamma$-rays and/or other particles/waves preferentially in certain directions similar to the collective phenomena in relativistic heavy-ion collisions. It is also well known that the dimensionality plays an important role in simulating supernova explosions.  As indicated by Eq.\,(\ref{def_pp1}), the 
ratio $E^{\rm{kin}}_{\rm{sym,4}}(\rho)/E^{\rm{kin}}_{\rm{sym}}(\rho)$ is zero in both 1D and 2D (in fact the quartic terms in 1D and 2D are identically zero), while it is $1/27$ in 3D as normally used in preparing the EOS for modeling neutron stars and various studies of heavy-ion reactions. Future studies of nuclear EOSs in 1D and 2D and their observational effects in collectively moving sub-systems in heavy-ion reactions and/or neutron stars might be interesting. 

\subsection{Relativistic FFG Model Predictions in 3D}
The relativistic kinetic energy per nucleon in ANM (in 3D) is obtained straightforwardly from
\begin{align}
\label{def_EOSANMReFFG}
E^{\rm{kin}}(\rho,\delta)=&\left.\left[\sum_{J=\rm{n,p}}\int_0^{k_{\rm{F}}^{J}}\d\v{k}
\left[\sqrt{\v{k}^2+M^2}-M\right]\right]\right/2\int_0^{k_{\rm{F}}}\d\v{k},
\end{align}
by generalizing the nucleon's dispersion relation from $\v{k}^2/2M$ to $\sqrt{\v{k}^2+M^2}-M$.
We recall that for small $\delta x$, one has mathematically 
\begin{align}
&\sum_{\Delta=\pm1}\rm{arcsinh}\,(x+\Delta\delta x)\approx2\rm{arcsinh}\,x
-\frac{x}{(x^2+1)^{3/2}}\delta x^2\notag\\
&-\frac{1}{4}\frac{x(2x^2-3)}{(x^2+1)^{7/2}}\delta x^4-\frac{1}{24}\frac{x(8x^4-40x^2+15)}{(x^2+1)^{11/2}}\delta x^6,
\end{align}
and  $\sqrt{1+\delta x}\approx1+2^{-1}\delta x-8^{-1}\delta x^2+16^{-1}\delta x^3-(5/128)\delta x^4+(7/256)\delta x^5-(21/1024)\delta x^6$. 
Using these approximations, one can obtain the following analytical expressions for the components of the $E^{\rm{kin}}(\rho,\delta)$ when it is expanded as an even-power series of $\delta$
\begin{align}
E_0^{\rm{kin}}(\rho)=&\frac{3M}{8\nu^2}\left[\left(1+2\nu^2\right)\left({1+\nu^2}\right)^{1/2}-\frac{\rm{arcsinh}\,\nu}{\nu}\right]-M,\\
E_{\rm{sym}}^{\rm{kin}}(\rho)=&\frac{M}{6}\frac{\nu^2}{\sqrt{1+\nu^2}},\\
E_{\rm{sym,4}}^{\rm{kin}}(\rho)=&
\frac{M}{648}\frac{\nu^2(10\nu^4+11\nu^2+4)}{(1+\nu^2)^{5/2}},\\
E_{\rm{sym,6}}^{\rm{kin}}(\rho)=&\frac{M}{34992}\frac{\nu^2}{(1+\nu^2)^{9/2}}\notag\\
&\times\left(176\nu^8+428\nu^6+477\nu^4+260\nu^2+56\right),
\end{align}
where the dimensionless quantity $\nu$ is defined as
$
\nu={k_{\rm{F}}}/{M}.
$

At the non-relativistic limit with $\nu\ll1$, the above expressions are reduced to
\begin{align}
E_0^{\rm{kin}}(\rho)\approx&\frac{3k_{\rm{F}}^2}{10M}\left(1-\frac{5}{28}\nu^2
+\frac{5}{72}\nu^4-\frac{25}{704}\nu^6\right),\label{non0}\\
E^{\rm{kin}}_{\rm{sym}}(\rho)\approx&\frac{k_{\rm{F}}^2}{6M}\left(1-\frac{1}{2}\nu^2+\frac{3}{8}\nu^4-\frac{5}{16}\nu^6\right),\label{non2}\\
E^{\rm{kin}}_{\rm{sym,4}}(\rho)\approx &\frac{k_{\rm{F}}^2}{162M}\left(1+\frac{1}{4}\nu^2-\frac{25}{32}\nu^6\right),\label{non4}\\
E^{\rm{kin}}_{\rm{sym,6}}(\rho)\approx &\frac{7k_{\rm{F}}^2}{4374M}\left(1+\frac{1}{7}\nu^2-\frac{5}{112}\nu^6\right),\label{non6}
\end{align}
where the in-front coefficient of each expression is the non-relativistic correspondence.
It is necessary to point out that the first-order relativistic correction to the kinetic symmetry energy, i.e., $-2^{-1}\nu^2\cdot(k_{\rm{F}}^2/6M)=-12^{-1}k_{\rm{F}}^4/M^3$, was first given in Ref.\,\cite{Fri05}, and its value at $\rho\approx0.16\,\rm{fm}^{-3}$ is found to be about $-$0.48\,MeV.

On the other hand, at the ultra-relativistic limit with $\nu\gg1$, we can similarly obtain ($\mu=\nu^{-1}$) the following
\begin{align}
E^{\rm{kin}}_0(\rho)\approx&\frac{3k_{\rm{F}}}{4}\left(1-\frac{4\mu}{3}+\mu^2
+\frac{1-4\ln(2/\mu)}{8}\mu^4\right),\label{rr0}\\
E^{\rm{kin}}_{\rm{sym}}(\rho)\approx&\frac{k_{\rm{F}}}{6}\left(1-\frac{1}{2}\mu^2+\frac{3}{8}\mu^4\right),\label{rr2}\\
E^{\rm{kin}}_{\rm{sym,4}}(\rho)\approx&\frac{5k_{\rm{F}}}{324}\left(1-\frac{7}{5}\mu^2+\frac{81}{40}\mu^4\right),\label{rr4}\\
E^{\rm{kin}}_{\rm{sym,6}}(\rho)\approx&\frac{11k_{\rm{F}}}{2187}\left(1-\frac{91}{44}\mu^2+\frac{729}{176}\mu^4\right),\label{rr6}
\end{align}
where each in-front coefficient is the corresponding expression in the ultra-relativistic case.

The density dependence of the above quantities, i.e., (\ref{non0}) to (\ref{rr6}), is rather different for small and large densities.
Take the symmetry energy as an example, we find in the non-relativistic limit that $E^{\rm{kin}}_{\rm{sym}}(\rho)\approx k_{\rm{F}}^2/6M\sim k_{\rm{F}}\nu\sim\rho^{2/3}\ll k_{\rm{F}}$ since $\nu\ll1$. While in the ultra-relativistic limit, $E_{\rm{sym}}^{\rm{kin}}(\rho)\approx k_{\rm{F}}/6\sim M\nu\sim\rho^{1/3}\gg M$ since now $\nu\gg1$.
All these quantities can take any positive values, while their ratios are bounded.

Now considering the ratio between the fourth-order and second-order terms, we immediately obtain
\begin{equation}\label{Psi_nu}
\Psi=\frac{E_{\rm{sym,4}}^{\rm{kin}}(\rho)}{E_{\rm{sym}}^{\rm{kin}}(\rho)}
=\frac{1}{108}\frac{4+11\nu^2+10\nu^4}{1+2\nu^2+\nu^4}.
\end{equation}
In the non-relativistic limit, the above expression is reduced to $\Psi\approx1/27+\nu^2/36-\nu^6/36+\mathcal{O}(\nu^8)\approx \Psi_{\rm{NR}}=1/27\approx 3.7\%$,  which is the familiar
result obtained earlier (in subsection \ref{SEC_II_A}).
On the other hand, we have in the ultra-relativistic limit that $\Psi\to\Psi_{\rm{UR}}=5/54\approx9.3$\% by taking $\nu\to\infty$ in (\ref{Psi_nu}), since now $\Psi\approx5/54-1/12\nu^2+1/9\nu^4-5/36\nu^6+\mathcal{O}(\nu^{-8})$.
Thus, between the two limits we have $1/27\leq \Psi\leq 5/54$. 
The enhancement from $\Psi_{\rm{NR}}=1/27$ to $\Psi_{\rm{UR}}=5/54$ is $5/54-1/27=1/18$ which is $150$\% (1.5 times) of the non-relativistic value $\Psi_{\rm{NR}}=1/27$, although the enhancement itself (i.e., 1/18) is still a small quantity. It is also interesting to point out that the Fermi momentum $k_{\rm{F}}$ is the only scale in the ultra-relativistic situation, since here the static mass $M\ll k_{\rm{F}}$, and thus could be neglected safely.
In particular, we have $E_{\rm{sym}}^{\rm{kin}}(\rho)\approx k_{\rm{F}}/6$ and $E_{\rm{sym,4}}^{\rm{kin}}(\rho)\approx5k_{\rm{F}}/324$.
In this sense, the relativistic effects enhance the quartic contribution (compared with the conventional quadratic kinetic symmetry energy).
Moreover, if one takes $\nu=k_{\rm{F}}/M\approx1$ roughly as the boundary between the non-relativistic and the relativistic regions, then $\rho\approx45\rho_0$ is found. It indicates that for conventional nuclear physics even for neutron stars ($\rho\lesssim10\rho_0$), the smallness of the quartic kinetic symmetry energy could be partially explained as due to the fact that the problems encountered in these studies are effectively non-relativistic.

 \renewcommand*\figurename{\small Fig.}
\begin{figure}[h!]
\centering
\includegraphics[height=3.8cm]{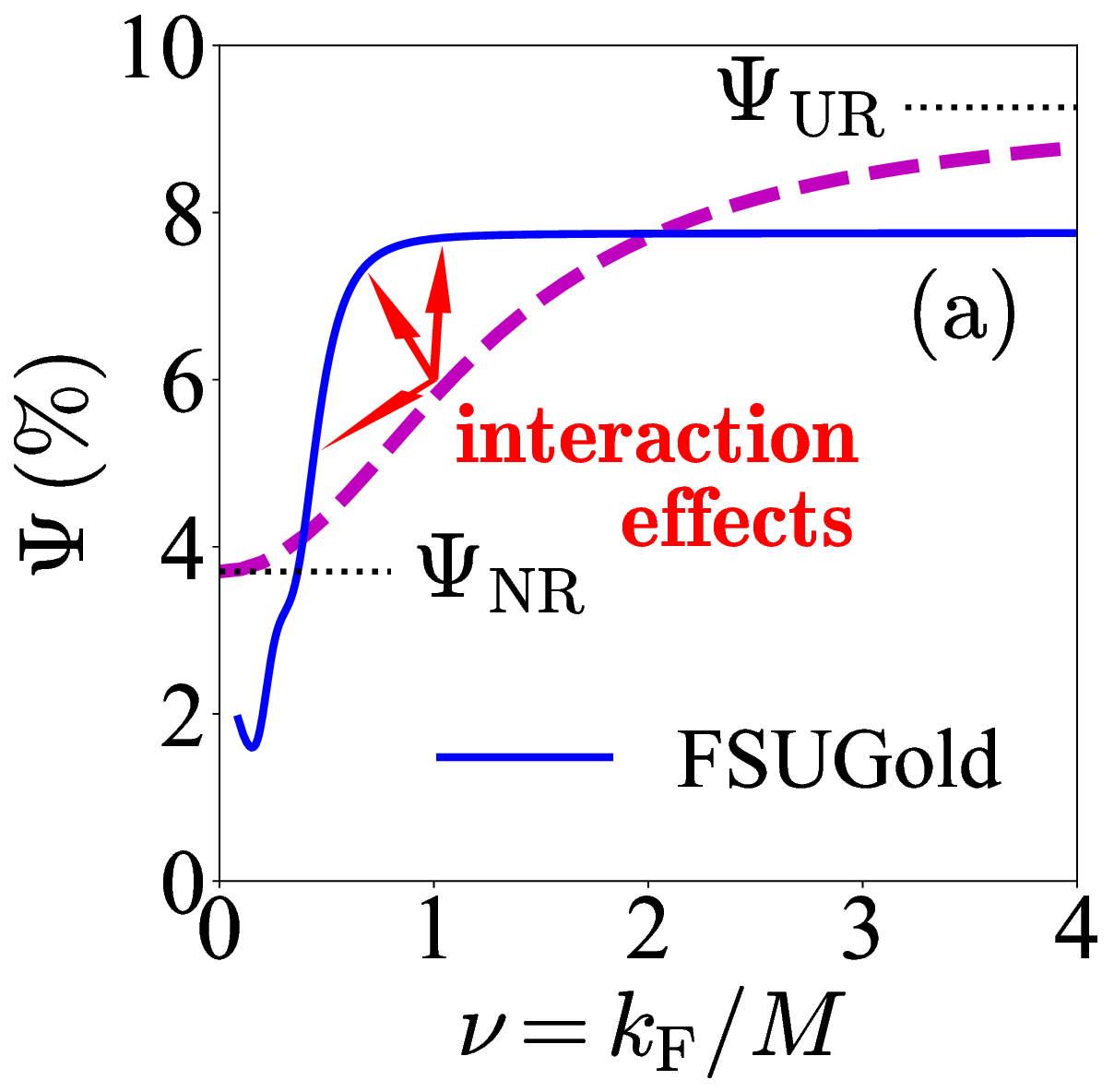}\quad
\includegraphics[height=3.8cm]{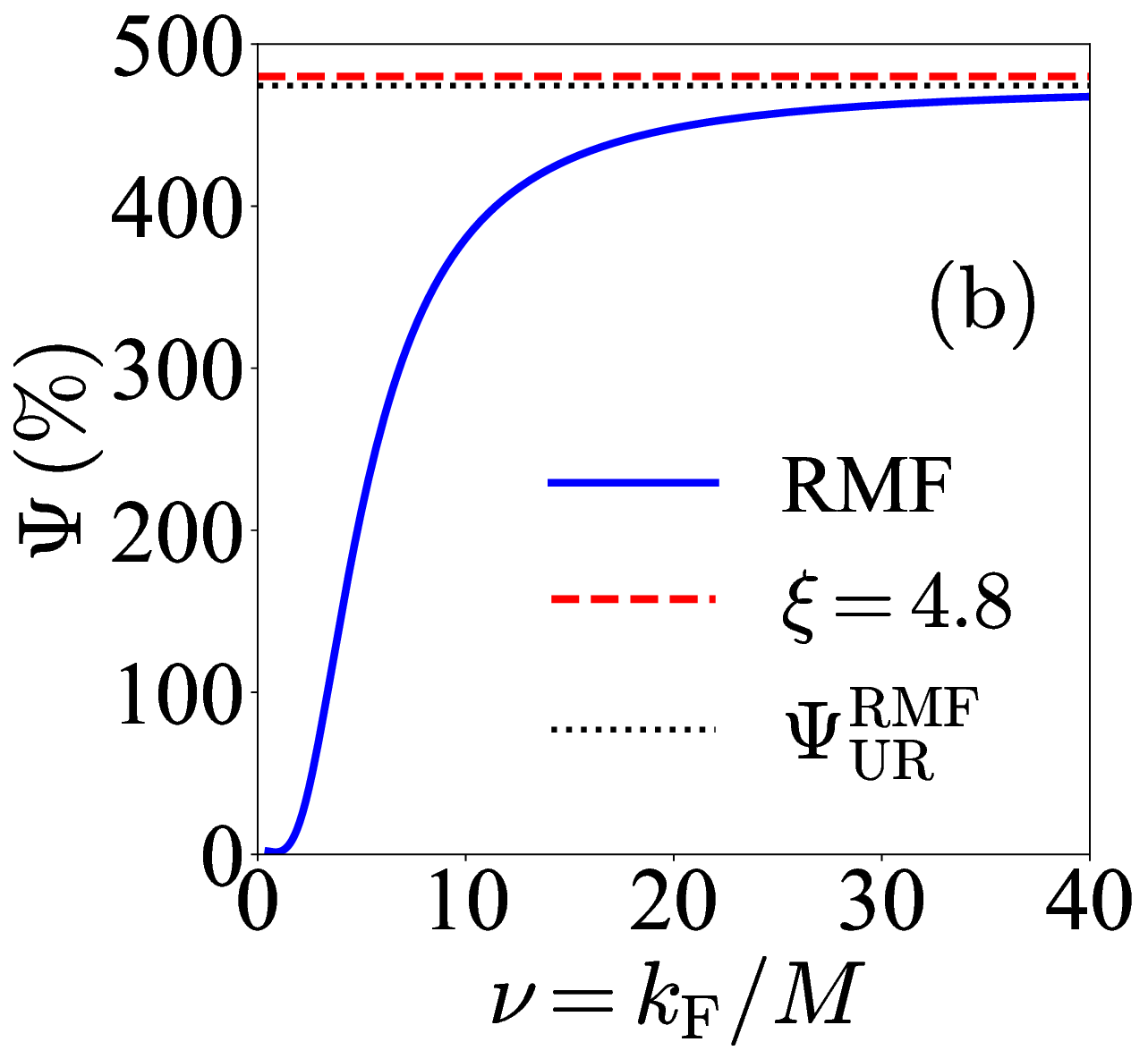}
\caption{(Color Online). Panel (a): The $\nu$ (equivalently $\rho$) dependence of the ratio $\Psi$ in the relativistic FFG model (red) in comparison with the nonlinear RMF model prediction (blue lines). 
Panel (b): results with a general RMF construction different from the FSUGold set, see detailed descriptions in the text.}\label{fig_Psi_nu}
\end{figure}
\section{$E_{\textmd{sym},4}(\rho)/E_{\textmd{sym}}(\rho)$ in Nonlinear Relativistic Mean Field Models: Effects of Nuclear Interactions}\label{SEC_III}
In order to investigate effects of nuclear effective interactions on the ratio $\Psi$, we give an example in this section adopting the nonlinear relativistic mean field (RMF) model.  In particular, the $\Psi$ obtained from the RMF using the FSUGold parameter set\,\cite{Tod05} is shown in the left panel of Fig.\,\ref{fig_Psi_nu} with the blue solid line,
while the dashed magenta line is the relativistic FFG model prediction according to Eq.\,(\ref{Psi_nu}).
It is clearly shown that the interaction changes significantly the FFG model prediction for $\Psi$ (indicated by the red arrows in panel (a)), at both low and high densities.
The total quadratic and quartic symmetry energy in the nonlinear RMF model at very high densities (ultra-relativistic limit) are,
\begin{align}
E_{\rm{sym}}(\rho)\approx& \frac{k_{\rm{F}}}{6}+\frac{c_{\omega}^{2/3}\rho^{1/3}}{2\Lambda_{\rm{V}}},\\
E_{\rm{sym,4}}(\rho)\approx&\frac{5k_{\rm{F}}}{324}+\frac{c_{\omega}^{5/3}\rho^{1/3}}{6\Lambda_{\rm{V}}^2},
\end{align}
respectively, and they both scale as $\rho^{1/3}$\,\cite{Cai12}.
Here the coupling parameters $c_{\omega}$ and $\Lambda_{\rm{V}}$ characterize the self-interaction among the four $\omega$ mesons as well as the coupling between the $\omega$ meson and the $\rho$ meson, respectively.
They are introduced into the nonlinear RMF Lagrangian $\mathcal{L}_{\rm{non-RMF}}$ through the $4^{-1}c_{\omega}g_{\omega}^4(\omega_{\mu}\omega^{\mu})^2$ and $2^{-1}\Lambda_{\rm{V}}g_{\rho}^2g_{\omega}^2\omega_{\mu}\omega^{\mu}\vec{\rho}_{\nu}\cdot\vec{\rho}^{\nu}$ terms, where $g_{\omega}$ and $g_{\rho}$ are two coupling constants between nucleons and the vector mesons\,\cite{Cai12}, respectively.
Consequently, we obtain the ratio of the quartic over the quadratic symmetry energy in the ultra-relativistic limit as
\begin{equation}\label{rmf-case}
\Psi=\Psi^{\rm{RMF}}_{\rm{UR}}\equiv\left.\left(\frac{5a}{324}+\frac{c_{\omega}^{5/3}}{6\Lambda_{\rm{V}}^2}\right)
\right/\left(\frac{a}{6}+\frac{c_{\omega}^{2/3}}{2\Lambda_{\rm{V}}}\right),
\end{equation}
where $a=\sqrt[3]{3\pi^2/2}$, see, e.g., Ref.\,\cite{Cai12} for more details on these analytical expressions.
In the FSUGold parameter set, $c_{\omega}=0.01$ and $\Lambda_{\rm{V}}=0.24$, and consequently $\Psi_{\rm{UR}}^{\rm{RMF}}\approx7.8$\%. As indicated in the panel (a), this limit is reached quickly as soon as $\nu=k_F/M$ becomes larger than about 1. We notice that in both the numerator and denumerator of the above expression, the first term is the kinetic while the second one is the potential contribution. The potential/kinetic ratio is
$(c_{\omega}^{2/3}/2\Lambda_{\rm{V}})/(a/6)\approx0.24$ and $(c_{\omega}^{5/3}/6\Lambda_{\rm{V}}^2)/(5a/324)\approx0.04$, respectively, indicating that the kinetic part is dominant in the FSUGold parameter set for both the $E_{\rm{sym}}(\rho)$ and $E_{\rm{sym,4}}(\rho)$ at the ultra-relativistic limit. Moreover, if the coupling constant $c_{\omega}$ or $1/\Lambda_{\rm{V}}$ is very small (near zero), then the high-density limit of $\Psi$ naturally approaches $5/54$ as predicted by the relativistic FFG model, see expression (\ref{Psi_nu}).

To further explore effects of nuclear interactions on the ratio $\Psi$, we now go beyond the FSUGold parameter set and check purely theoretically what will happen if the potential term 
$c_{\omega}^{5/3}/6\Lambda_{\rm{V}}^2$ is far larger than the kinetic term $5a/324$ and in the meanwhile $c_{\omega}^{2/3}/2\Lambda_{\rm{V}}$ is much larger than $a/6$. 
Obviously according to Eq.\,(\ref{rmf-case}), the $\Psi$ then approaches another constant $\xi=c_{\omega}/3\Lambda_{\rm{V}}$ independent of the density $\rho$ by neglecting the kinetic contributions.
In this case, the fourth-order symmetry energy is not necessarily smaller than the quadratic one, i.e., the $E_{\rm{sym},4}(\rho)$ could be comparable or even be (much) larger than the $E_{\rm{sym}}(\rho)$ if the density is large enough. In other words, nucleon-nucleon interaction could change significantly the relative strength of the quadratic and quartic symmetry energies at large densities. 
Numerically, the ratio $\Psi$ could even approach infinity if the coupling constant $\Lambda_{\rm{V}}$ is selected to be very small (but not zero). Of course, this may only happen at extremely high densities where there is no experimental and/or theoretical constraints and may not be reachable anywhere. Thus, this exercise may be only for satisfying our intellectual curiosity. 

In order to check the above expectations, we show in the right panel of Fig.\,\ref{fig_Psi_nu} an example of a relevant RMF construction for $\Psi$ (blue solid line), the $\Psi_{\rm{UR}}^{\rm{RMF}}$ factor (black dotted line) by
setting artifically the constant $\xi=c_{\omega}/3\Lambda_{\rm{V}}=4.8$. Here the coupling constant $c_{\omega}$ is still fixed at 0.01 as in the FSUGold, but the coupling constant $\Lambda_{\rm{V}}$ is adjusted according to $\Lambda_{\rm{V}}=c_{\omega}/3\xi\approx7\times10^{-4}$ (which is much smaller than the one used in the FSUGold set). Thus, overall, the newly constructed RMF parameter set is different from the FSUGold one.
It is seen that as the dimensionless quantity $\nu$ becomes large, the RMF prediction naturally approaches $\Psi_{\rm{UR}}^{\rm{RMF}}$. It also approaches the constant factor $\xi$ since the interaction parts in both the quadratic and quartic symmetry energies are dominant over their kinetic parts. More quantitatively, under these conditions for $c_{\omega}$ and $\Lambda_{\rm{V}}$ we have $c_{\omega}^{2/3}/2\Lambda_{\rm{V}}\approx33.42$, which is far larger than the kinetic term $a/6\approx0.41$, while similarly $c_{\omega}^{5/3}/6\Lambda_{\rm{V}}^2\approx160.41$ which is also far larger than the kinetic contribution $5a/324\approx0.04$. It is necessary to point out that a factor $\nu$ about 40 corresponds to a density $\rho$ about $2.9\times10^6\rho_0$, which is thus only theoretically meaningful. 
Since the large-$\rho$ limit is also the relativistic limit,  we find that the combination of the nucleon-nucleon interactions and the relativistic corrections (here come into play in the potential part) essentially modify the prediction on the ratio $\Psi$, although the relativistic corrections alone can not effectively affect the value of $\Psi$.
In this sense, the nucleon-nucleon interactions (here through the effective potentials in the EOS of ANM) are very important for the ratio $\Psi$ as one generally expects.

On the opposite side, i.e., in the ultra-low density limit,  the kinetic symmetry energy from the nonlinear RMF models could be expanded as $E^{\rm{kin}}_{\rm{sym}}(\rho)\approx (k_{\rm{F}}^2/6M)\cdot[1-2^{-1}(k_{\rm{F}}/M)^2+M^{-1}(g_{\sigma}/m_{\sigma})^2\rho]+\cdots\sim b\nu^2+\cdots$, while the potential part is similarly approximated as $E_{\rm{sym}}^{\rm{pot}}(\rho)\approx2^{-1}(g_{\rho}/m_{\rho})^2\rho\cdot[1-\Lambda_{\rm{V}}(g_{\rho}/m_{\rho})^2(g_{\omega}/m_{\omega})^4\rho^2]+\cdots\sim b'\nu^3+\cdots$, where $b$ and $b'$ are two constants, $g_{\sigma}$ is the coupling constant between the nucleon and the $\sigma$ meson, and $m_{\sigma},m_{\rho}$ and $m_{\omega}$ are the static masses of the $\sigma,\rho$ and the $\omega$ mesons, respectively. Thus the ratio $E_{\rm{sym}}^{\rm{pot}}(\rho)/E_{\rm{sym}}^{\rm{kin}}(\rho)\approx (b'/b)\nu$ approaches zero as $\rho\to0$. One can find similarly that at ultra-low densities, the fourth-order symmetry energy is dominated by its kinetic part, actually according to the non-relativistic FFG prediction. More quantitatively, we have $E_{\rm{sym},4}^{\rm{kin}}(\rho)\approx k_{\rm{F}}^2/162M+\cdots\sim c\nu^2+\cdots$, and similarly $E_{\rm{sym},4}^{\rm{pot}}(\rho)\approx-72^{-1}(g_{\sigma}/m_{\sigma})^2\rho\nu^4+\cdots\sim c'\nu^7+\cdots$ by carefully considering the exact expression for the fourth-order symmetry energy\,\cite{Cai12}, where $c$ and $c'$ are another two constants.
Consequently, the ratio $E_{\rm{sym},4}^{\rm{pot}}(\rho)/E_{\rm{sym},4}^{\rm{kin}}(\rho)\approx (c'/c)\nu^5$ which is much smaller than 1 since $\nu\ll1$ for low densities.
From these low-density expansions, we can conclude that although the nucleon-nucleon interactions could modify the ratio $\Psi$,  it is still mainly determined by the kinetic contributions at low densities.

\section{Relativistic Corrections to Kinetic Symmetry Energies in the Presence of Short-Range Correlations}\label{S4}

\subsection{Nucleon Momentum Distribution with SRC-induced High-Momentum Nucleons}\label{S3}
In this subsection, we briefly recall the single nucleon momentum distribution function encapsulating a SRC-induced HMT and the relevant parameters.
Based on predictions of microscopic nuclear many-body theories and relevant experimental
findings\,\cite{Hen14,Due18,Sch19,Sch20}, the single-nucleon momentum distribution in ANM can be parametrized as\,\cite{Cai16},
\begin{equation}\label{MDGen}
n^J_{\v{k}}(\rho,\delta)=\left\{\begin{array}{ll}
\Delta_J,~~&0<|\v{k}|<k_{\rm{F}}^J,\\
\displaystyle{C}_J\left({k_{\rm{F}}^{J}}/{|\v{k}|}\right)^4,~~&k_{\rm{F}}^J<|\v{k}|<\phi_Jk_{\rm{F}}^J.
\end{array}\right.
\end{equation}
Here, $\Delta_J$ is the depletion of the Fermi sphere with respect to the step function in the FFG model.
The three parameters $\Delta_J$, ${C}_J$ and $\phi_J$ are constrained by the
fraction of nucleons in the HMT 
\begin{equation}\label{def_xJHMT}
x_J^{\rm{HMT}}=\left.\int_{k_{\rm{F}}^J}^{\phi_Jk_{\rm{F}}^J}
n_{\v{k}}^J\d\v{k}\right/{\displaystyle\int_0^{\phi_Jk_{\rm{F}}^J}
n_{\v{k}}^J\d\v{k}}=3C_{{J}}\left(1-\frac{1}{\phi_{{J}}}\right),
\end{equation} and the normalization condition 
\begin{equation}\label{def_NC}
\frac{2}{(2\pi)^3}\int_0^{\infty}n^J_{\v{k}}(\rho,\delta)\d\v{k}=\rho_J=(k_{\rm{F}}^{J})^3/3\pi^2.
\end{equation} 
Only two of the three parameters $\Delta_J$, $C_J$ and
$\phi_J$ are independent. Using the last two as independent parameters and assuming they have the same isospin dependence, i.e, 
$Y_J=Y_0(1+Y_1\tau_3^J\delta)$\,\cite{Cai16}, the associated parameters were then constrained to be about $C_0\approx0.161\pm0.015,C_1\approx-0.25\pm0.07,\phi_0\approx2.38\pm0.56$ and $\phi_1\approx-0.56\pm0.10$ using information from the SRC experiments\,\cite{Hen14,Hen16x}, see, e.g., Ref.\,\cite{LCCX18} for a recent review. 
 \renewcommand*\figurename{\small Fig.}
\begin{figure*}
\centering
\includegraphics[height=3.7cm]{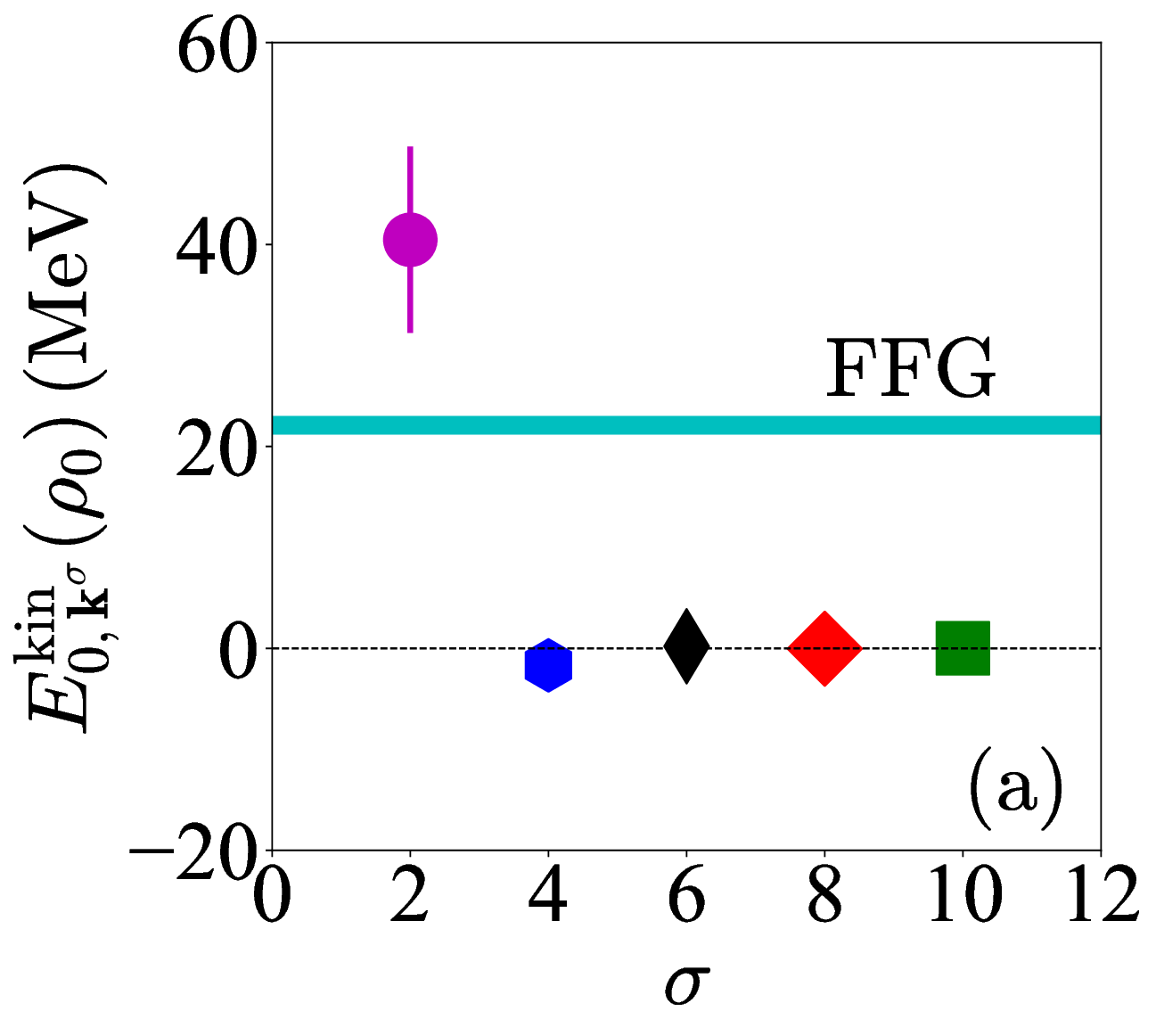}\quad
\includegraphics[height=3.7cm]{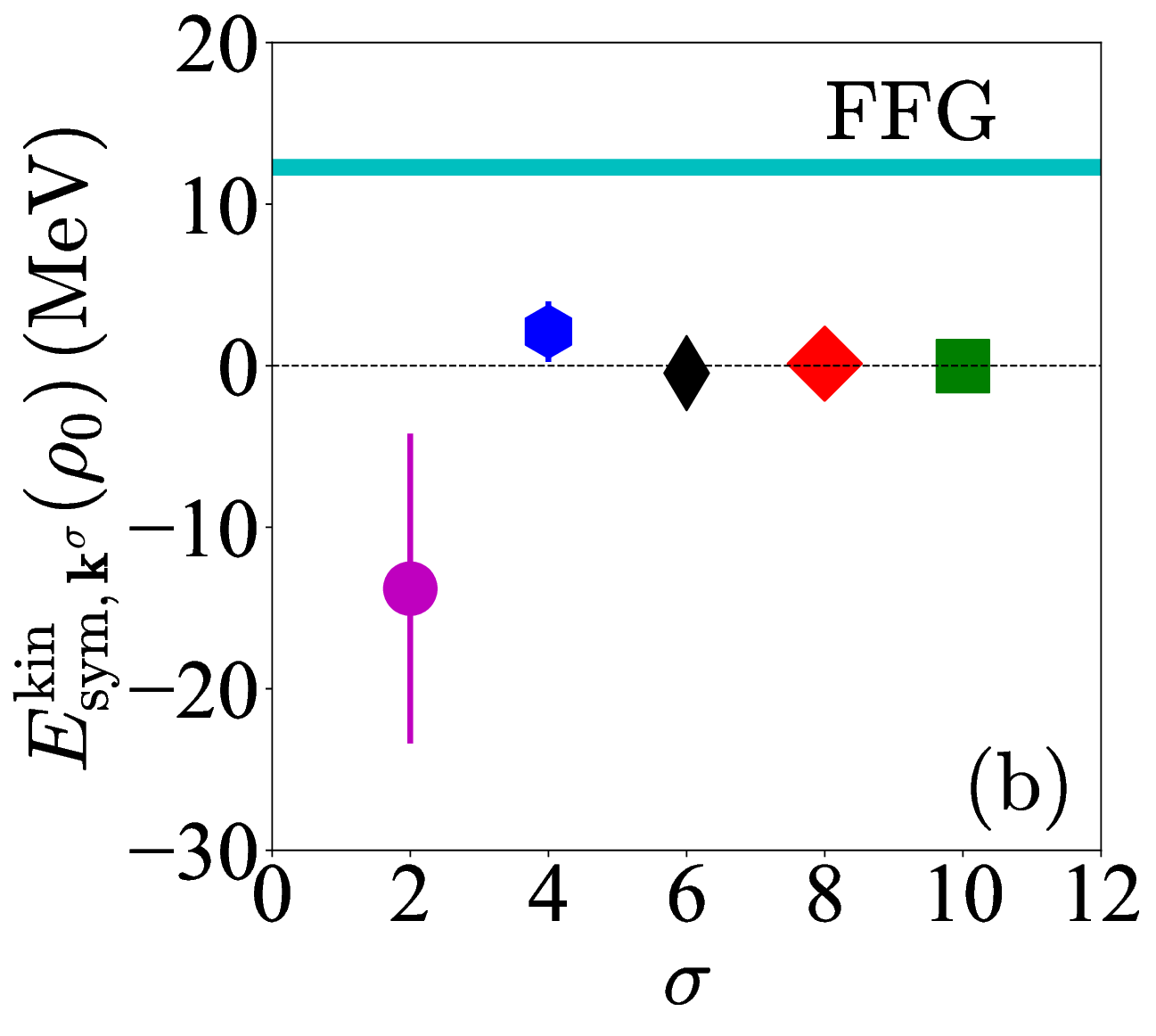}\quad
\includegraphics[height=3.7cm]{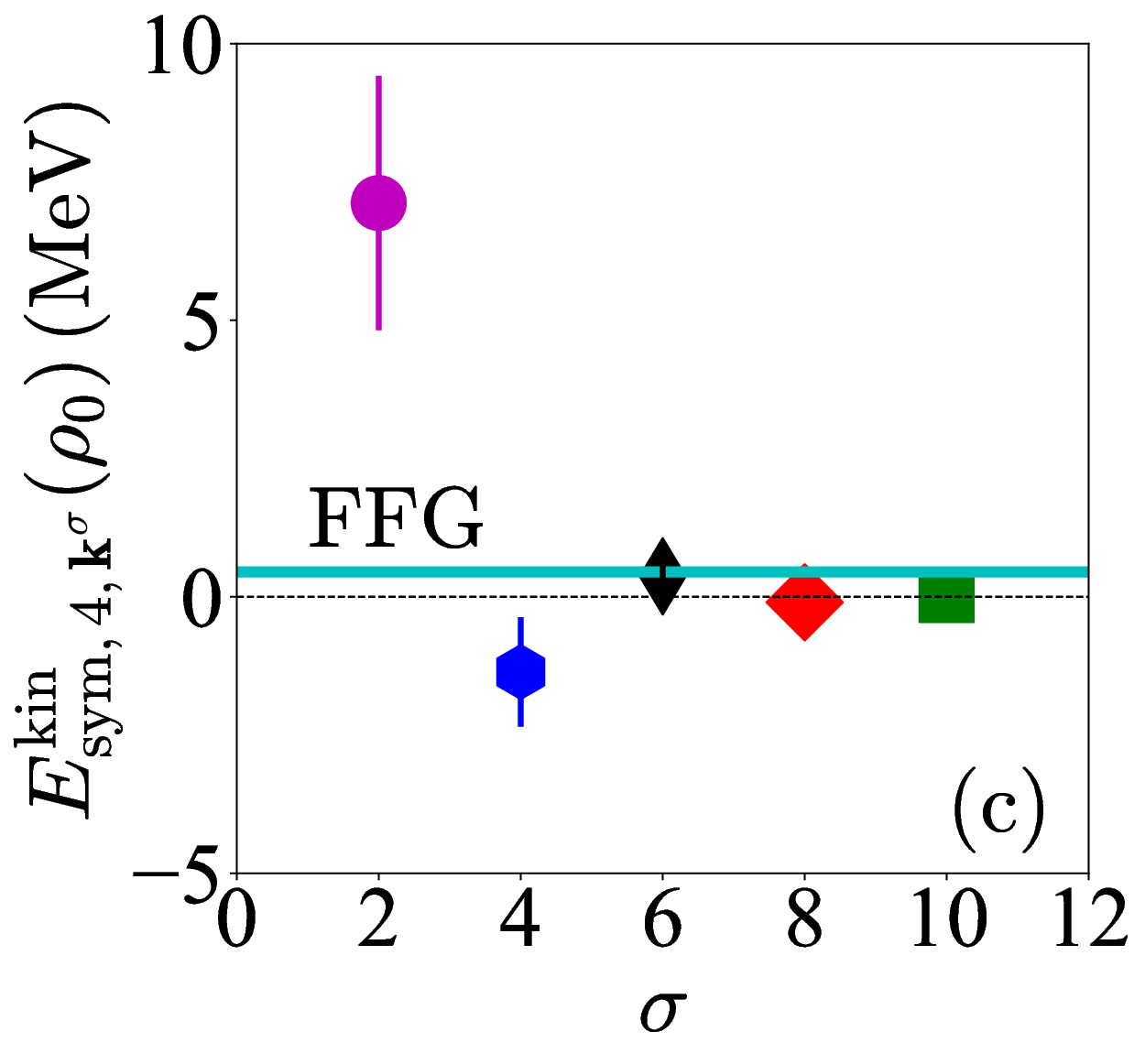}\quad
\includegraphics[height=3.7cm]{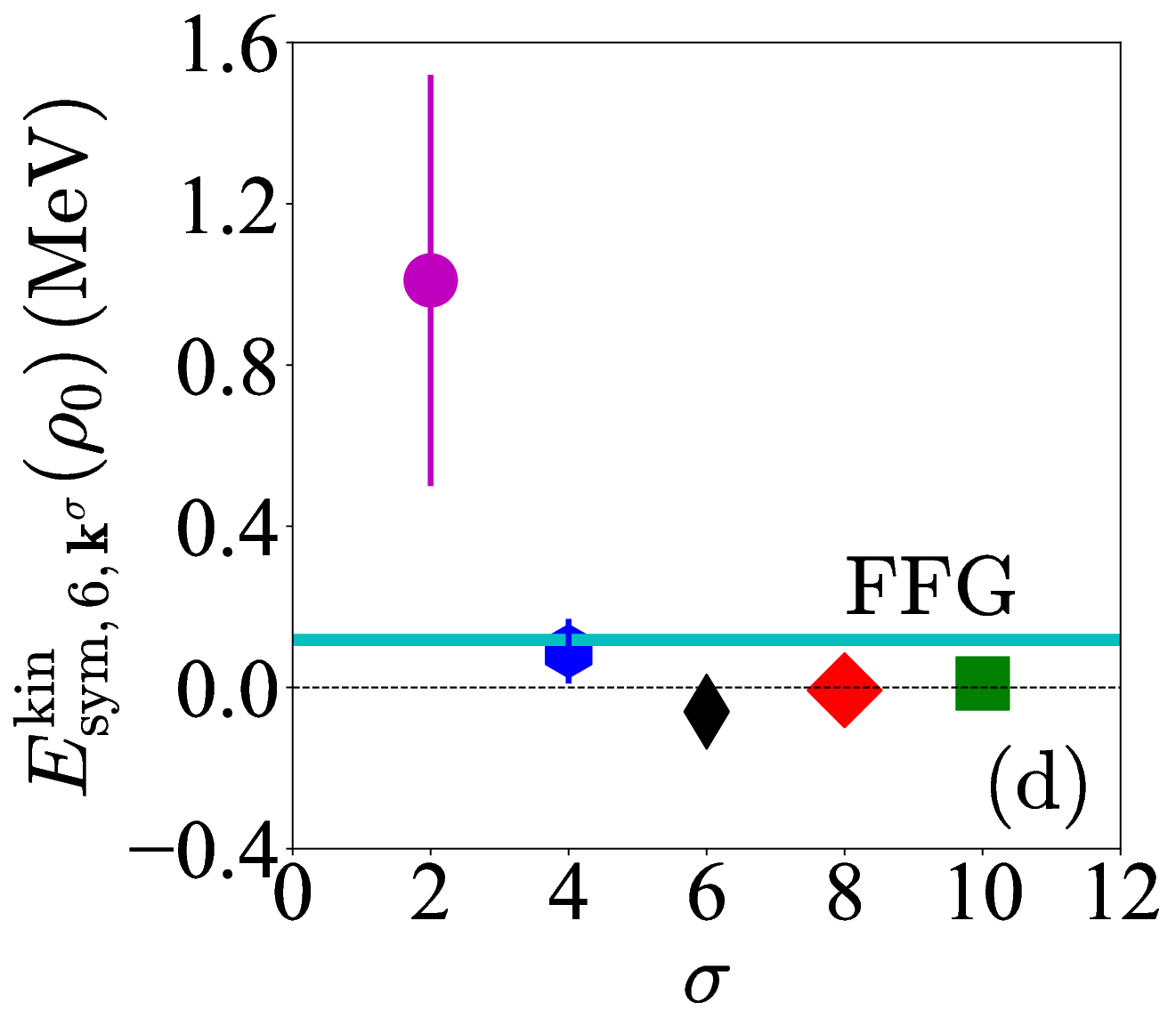}
 \caption{(Color Online).  Kinetic parts of the EOS of ANM with its relativistic corrections for $E_0^{\rm{kin}}(\rho_0),E_{\rm{sym}}^{\rm{kin}}(\rho_0),E_{\rm{sym,4}}^{\rm{kin}}(\rho_0)$ and $E_{\rm{sym,6}}^{\rm{kin}}(\rho_0)$.
 Here the saturation density $\rho_0\approx0.16\pm0.01\,\rm{fm}^{-3}$, $\sigma=2$ is the non-relativistic term while $\sigma=4,6,8$ and $10$ correspond to the first four relativistic corrections to the kinetic parts of the EOS originated from Eq.\,(\ref{def_nonexp}).}\label{fig_RC}
\end{figure*}

The isospin dependence of the SRC-induced HMT was found to affect especially the kinetic part of nuclear symmetry energy\,\cite{Xulili,Hen15,Car12,Vid11,Lov11}. It has also been found to affect significantly properties of neutron stars, such as the mass-radius relation, tidal polarizability, cooling rate and crust-core transition density\,\cite{Cai16a,Sou20-a,Sou20-b,HLu22}. Interestingly, the enhanced pressure due to the SRC-induced high momentum nucleons can also help balance the reduction of the maximum mass caused by the possible existence of dark matter particles in neutron stars\,\cite{Lou22}. There is also a long history of studying effects of SRC on nuclear reactions\,\cite{Ant93}. More recently, the SRC-induced HMT has been found to affect particularly the dynamics and emissions of energetic nucleons, hard photons and/or the production of deeply sub-threshold particles in nuclear reactions\,\cite{LiBA15,Yong17-a,Yong17-b,Wang17,Guo21}.  Interestingly, a very recently analysis of the energy spectra of protons emitted in reactions of 47\,MeV/u projectiles with
Sn and Au targets provides new evidence for the $1/k^4$ shape of the HMTs in the intrinsic momenta spectra of the projectiles\,\cite{Chris}.

\subsection{First Four Relativistic Corrections to the Kinetic EOS of ANM Considering the SRC-induced HMT}

As shown in Eq.\,(\ref{def_nonexp}), the first four relativistic corrections to the nucleon kinetic energy are given by 
``$-\v{k}^2/4M^2$'', ``$\v{k}^4/8M^4$'', ``$-5\v{k}^6/64M^6$'' and ``$7\v{k}^8/128M^8$'', respectively.  In calculating the average energy per nucleon in ANM, after integrating over the momentum $k$, the dependence of the kinetic energy on momentum $k$ is transformed into the corresponding dependence on the Fermi momentum $k_{\rm{F}}\sim\rho^{1/3}$. 
Specifically, for a general $\sigma$ the term $\v{k}^{\sigma}/M^{\sigma-1}$ in the kinetic energy leads to the $3k_{\rm{F}}^{\sigma}/(\sigma+3)M^{\sigma-1}$ term in the kinetic EOS if the step function for the momentum distribution $n_{\v{k}}^J$ is adopted. Similarly, the corresponding term for the kinetic symmetry energy is $\sigma k_{\rm{F}}^{\sigma}/6M^{\sigma-1}$.
For example, for the first-order relativistic correction to the kinetic energy, one has the relevant corrections for the kinetic EOS of SNM and for the kinetic symmetry energy as $-3k_{\rm{F}}^4/56M^3$ and $-k_{\rm{F}}^4/12M^3$\,\cite{Fri05}, respectively, see also the expressions (\ref{non0}) and (\ref{non2}).

However, when the SRC-induced HMT is considered,  the analytical expression for $\langle \v{k}^{\sigma}\rangle$ becomes non-trivial.
The $\langle \v{k}^\sigma\rangle$ with a general power $\sigma$ at order $\delta^0$ (corresponding to SNM), $\delta^2$ (corresponding to the quadratic symmetry energy) and $\delta^4$ (corresponding to the quartic symmetry energy) are given in the appendix in details.
For our interest here,  terms in the expansion (\ref{def_nonexp}) will be investigated numerically below using $\sigma=2,4,6,8$ and $10$.
In addition, a similar expression for the $\langle\v{k}^{\sigma}\rangle$ at order $\delta^6$, which is relevant for the derivation of the sixth-order symmetry energy, could also be obtained similarly.
While we do not give its explicit expression in the appendix due to its very complicated form, numerical results for the sixth-order symmetry energy will also be presented below.

Using the parameters $\phi_0,\phi_1,C_0$ and $C_1$ given in the last subsection, we can obtain the magnitudes of all terms in the kinetic EOS as well as their relativistic corrections considering the SRC effects.
In particular, we have found that $E_{0,\v{k}^2}^{\rm{kin}}(\rho_0)\approx40.47\pm9.23\,\rm{MeV}$ adopting $\rho_0\approx0.16\pm0.01\,\rm{fm}^{-3}$, here the subscript ``$0,\v{k}^2$'' reminds us that this term originates from the non-relativistic kinetic energy $\v{k}^2/2M$, thus it is not a relativistic correction. To the next order in $\v{k}^{\sigma}$, we then have $E_{0,\v{k}^4}^{\rm{kin}}(\rho_0)\approx-1.68\pm1.11\,\rm{MeV}$, which originates from the first relativistic correction ``$-\v{k}^4/8M^3$'' (dividing the zeroth order term $\v{k}^2/2M$ gives ``$-\v{k}^2/4M^2$'' in Eq.\,(\ref{def_nonexp})).
Similarly, contributions from the next three terms of the relativistic corrections,  i.e., ``$\v{k}^6/16M^5$'', ``$-5\v{k}^8/128M^7$'', and ``$7\v{k}^{10}/256M^9$'', are found to be about $0.21\pm0.25\,\rm{MeV},-0.04\pm0.07\,\rm{MeV}$ and $0.01\pm0.02\,\rm{MeV}$, respectively.
It is obvious that the fourth-order relativistic correction with its value about $0.01\pm0.02\,\rm{MeV}$ is much smaller than the (leading) non-relativistic contribution, indicating that considering the first four relativistic corrections is enough for estimating the overall relativistic effects.

The magnitudes of the kinetic EOS of SNM together with its first four relativistic corrections are shown in the panel (a) of Fig.\,\ref{fig_RC}, where the non-relativistic FFG prediction about $22.10\pm0.92\,\rm{MeV}$ is also shown for a comparison (marked as ``FFG'').
The enhancement of the kinetic EOS of SNM considering the SRC-induced HMT is known for some time, see, e.g., Refs.\,\cite{Hen14,Xulili,Cai16a} and references therein.
By adding the non-relativistic kinetic EOS of SNM and its first four relativistic corrections, one then obtains,
\begin{equation}
E_0^{\rm{kin}}(\rho_0)\approx\sum_{\sigma=2\ell}^{\ell=1\sim5}E_{0,\v{k}^\sigma}^{\rm{kin}}(\rho_0)\approx38.98\pm8.32\,\rm{MeV}.
\end{equation}
Compared with 40.47\,MeV,  a reduction about 1.49\,MeV on the (non-relativistic) $E_0^{\rm{kin}}(\rho_0)$ is generated due to the relativistic effects, and the relative correction is about 4\%.

Similarly, we can obtain the first four relativistic corrections to the quadratic kinetic symmetry energy as $2.11\pm1.87\,\rm{MeV},-0.46\pm0.58\,\rm{MeV},0.12\pm0.20\,\rm{MeV}$ and $-0.03\pm0.07\,\rm{MeV}$, respectively. The quadratic kinetic symmetry energy itself considering the SRC-induced HMT is found to be about $-13.80\pm9.57\,\rm{MeV}$. Based on these results, we have
\begin{equation}
E_{\rm{sym}}^{\rm{kin}}(\rho_0)\approx\sum_{\sigma=2\ell}^{\ell=1\sim5}E_{\rm{sym},\v{k}^\sigma}^{\rm{kin}}(\rho_0)\approx-12.01\pm8.23\,\rm{MeV}.
\end{equation}
Compared with the original quadratic kinetic symmetry energy encapsulating the HMT but no relativistic correction, i.e., $-13.80\pm9.57\,\rm{MeV}$, the relativistic corrections generate an enhancement about 1.79\,MeV, corresponding to a relative correction about 13\%, see the panel (b) of Fig.\,\ref{fig_RC}.
The value either $-13.80\,\rm{MeV}$ (without) or $-12.01\,\rm{MeV}$ (with) relativistic corrections for $E_{\rm{sym}}^{\rm{kin}}(\rho_0)$ is consistent with the value of $-16.94\,\rm{MeV}$
obtained directly from a nonlinear RMF model including the HMT\,\cite{Cai16a}. 

It is important to emphasize that the SRC-induced reduction of the quadratic kinetic symmetry energy with respect to the FFG prediction
of about $12.28\pm0.51\,\rm{MeV}$ at saturation density is also known for some time when the SRC effects are considered in microscopic nuclear many-body theories or simple phenomenlogical models, 
see, e.g., Ref.\,\cite{Carb14,LCCX18} for reviews. These studies include the Brueckner-Hartree-Fock (BHF) approach using the Av18 potential plus the Urbana IX three-body force\,\cite{Vid11},
the self-consistent Green's function (SCGF) theories adopting different microscopic interactions (N3LO, Av18, Nij1, and CD Bonn)\,\cite{Car12},   the Fermi
hypernetted chain (FHNC) method\,\cite{Lov11}, and phenomenological models\,\cite{Xulili,Hen15}. See the Fig.\,41 in Ref.\,\cite{LCCX18} for a summary of the results from some of these studies in comparison with the
values extracted from analyzing the SRC data within a neutron-proton dominance model\,\cite{Hen15}.
We also note that the quadratic kinetic symmetry energy was obtained approximately in some of the earlier studies by using the difference between
the EOSs of pure neutron matter (PNM) and SNM adopting the parabolic approximation of the ANM EOS.

Moreover, the kinetic fourth- and sixth-order symmetry energies considering the first four relativistic corrections in the presence of SRC are about $6.06\pm1.77\,\rm{MeV}$ and $1.07\pm0.54\,\rm{MeV}$, respectively, i.e.,
\begin{align}
E_{\rm{sym},4}^{\rm{kin}}(\rho_0)\approx\sum_{\sigma=2\ell}^{\ell=1\sim5}E_{\rm{sym,4},\v{k}^\sigma}^{\rm{kin}}(\rho_0)\approx6.06\pm1.77\,\rm{MeV},\\
E_{\rm{sym,6}}^{\rm{kin}}(\rho_0)\approx\sum_{\sigma=2\ell}^{\ell=1\sim5}E_{\rm{sym,6},\v{k}^\sigma}^{\rm{kin}}(\rho_0)\approx1.07\pm0.54\,\rm{MeV}.
\end{align}
More specifically, the contributing terms for the fourth-order symmetry energy $E_{\rm{sym},4}^{\rm{kin}}(\rho_0)$ are $7.12\pm2.30\,\rm{MeV}$ (non-relativistic contribution), $-1.36\pm0.99\,\rm{MeV}$, $0.37\pm0.44\,\rm{MeV}$, $-0.10\pm0.17\,\rm{MeV}$ and $0.03\pm0.06\,\rm{MeV}$,  respectively. Similarly, those for the sixth-order symmetry energy $E_{\rm{sym,6}}^{\rm{kin}}(\rho_0)$ are $1.01\pm0.51\,\rm{MeV}$ (non-relativistic contribution), $0.09\pm0.08\,\rm{MeV}$, $-0.06\pm0.07\,\rm{MeV}$, $-0.007\pm0.05\,\rm{MeV}$ and $0.01\pm0.05\,\rm{MeV}$, respectively. See the panels (c) and (d) of Fig.\,\ref{fig_RC}, respectively, for illustrations.
From these results, one finds that the relativistic corrections on the kinetic fourth- and sixth-order symmetry energies are about $-1.06\,\rm{MeV}$ and $0.06\,\rm{MeV}$, with the relative effects about 15\% and 6\%, respectively.
It is thus obvious that the absolute change on the sixth-order symmetry energy due to the relativistic corrections could be safely neglected compared with its non-relativistic value.

Finally, it is interesting to compare the ratio $\Psi$ between the kinetic quartic and quadratic symmetry energies with and without the relativistic corrections in the presence of SRC-induced high momentum nucleons. Based on the results presented above, the $\Psi$ changes from $\Psi\approx -7.12/13.80\approx-51.6\%$ (without) to $\Psi\approx-6.06/12.01\approx-50.4\%$ (with) the relativistic corrections. 
Moreover, considering the extreme case of PNM, the overall relativistic corrections can be evaluated by adding $E_0^{\rm{kin}}(\rho_0)$, $E_{\rm{sym}}^{\rm{kin}}(\rho_0)$, $E_{\rm{sym},4}^{\rm{kin}}(\rho_0)$ and $E_{\rm{sym},6}^{\rm{kin}}(\rho_0)$ alltogether. This sum is 34.8\,MeV for the non-relativistic FFG model and 34.1\,MeV by including the (first four) relativistic corrections, both are close to the non-relativistic FFG model prediction about $3k_{\rm{n}}^{2}/10M\approx35.1\,\rm{MeV}$ where $k_{\rm{n}}=(3\pi^2\rho)^{1/3}=2^{1/3}k_{\rm{F}}$ is the neutron Fermi momentum in the PNM. Thus, although the relativistic corrections to each kinetic energy term may be large or small, the overall effects are rather small (the relative effect is about 2\%), since some changes are positive while others are negative.
Particularly, for the four kinetic energies, only the (quadratic) symmetry energy is reduced considering the SRC-induced HMT, while the other three higher order symmetry energies are all enhanced, compared with their FFG predictions (i.e., without HMT), see Fig.\,\ref{fig_RC}. These results indicate clearly that the SRC-induced HMT is more fundamental and has much larger impacts on the ANM EOS than the relativistic corrections.
Although the SRC effects are introduced through the HMT in the single nucleon momentum distribution function $n_{\v{k}}^J$ and affect apparently only the kinetic parts of ANM EOS in the models considered here,  
the SRC is fundamentally due to the tensor force in the neutron-proton isosinglet interaction channel\,\cite{Hen14,Hen16x}. Thus, in this sense,  the finding of this section is consistent with that in section \ref{SEC_III}, i.e., the nucleon-nucleon interactions could change the value of $\Psi$ significantly.

\section{Summary}\label{S5}
In summary, to pin down the EOS of dense neutron-rich matter has long been a major science driver in both astrophysics and nuclear physics. Most studies within both non-relativistic and relativistic nuclear many-body theories and phenomenological models with and/or without considering the isospin dependence of SRC indicate that the $E(\rho,\delta)$ converges quickly when it is expanded in terms of even powers of $\delta$. Thus, the so-called empirical parabolic law of ANM EOS seems to be valid even as $\delta\rightarrow 1$. While there are many empirical evidences available in the literature using various many-body approaches and interactions, it has been unclear why the isospin quartic symmetry energy is so small compared with the quadratic one and if there is any deep physics reason for the seemingly quick convergence in expanding the $E(\rho,\delta)$. 

In this work, we tried to decipher effects of relativistic kinematics, dimensionality, interactions and SRC on the ratio $\Psi$ of isospin quartic over quadratic symmetry energies as transparent as possible.  Within both relativistic and non-relativistic FFG models in coordinate spaces of arbitrary dimension $d$ with and without considering SRC effects as well as the 3D RMF models at both low and high density limits, we learned a few things that might be useful for the community to further understand the EOS of dense neutron-rich nuclear matter. In particular, we found that the ratio $\Psi$ in the FFG model depends strongly on the dimension $d$. 
While the ratio $\Psi$ is very small already in the normal 3D space, it could be even smaller in spaces with reduced dimensions. Based on this finding and stimulated by the many interesting new physics found in two-component cold atoms moving in either the same or mixed 1D or 2D spaces, we pointed out a few situations where the EOS in spaces with reduced dimensions may be useful for heavy-ion reactions and/or neutron stars. 
We also found that the ratio $\Psi$ could theoretically become very large only at the ultra-relativistic limit far above the density reachable in neutron stars. On the other hand, the non-linear RMF mode using the FSUGold parameter set predicts a $\Psi$ value around $2-8\%$ and it has significantly different density dependence compared to the relativistic FFG model prediction.  In the relativistic FFG model incorporating the SRC-induced HMT, the SRC affects significantly not only the kinetic energy of SNM but also the ratio $\Psi$ while the relativistic corrections are negligible. In conclusion, while the relativistic kinematics and dimensionality may affect appreciably the ratio $\Psi$, it is the nuclear interactions and the associated SRC dominate the ratio $\Psi$. We also found no fundamental physics reason for the $\Psi$ to be very small especially at high densities. 

\section*{Acknowledgement}
This work is supported in part by the U.S. Department of Energy, Office of Science, under Award Number DE-SC0013702, the CUSTIPEN (China-U.S. Theory Institute for Physics with Exotic Nuclei) under the US Department of Energy Grant No. DE-SC0009971.
\appendix
\renewcommand\theequation{a\arabic{equation}}
\begin{widetext}
\section{Analytical Expressions for $\langle \v{k}^{\sigma}\rangle$ in Presence of SRC-induced HMT}
According to the definition of the average of $\v{k}^{\sigma}$, namely 
\begin{equation}
\langle|\v{k}|^{\sigma}\rangle=\langle\v{k}^{\sigma}\rangle=\left.\langle\v{k}^{\sigma}(\rho,\delta)\rangle=\sum_{J=\rm{n,p}}\int_0^{\phi_Jk_{\rm{F}}^J}\v{k}^\sigma n_{\v{k}}^J(\rho,\delta)\d\v{k}\right/2\int_0^{k_{\rm{F}}}\d\v{k}
,
\end{equation} one obtains by noticing that $2\int_0^{k_{\rm{F}}}\d\v{k}=\rho[2/(2\pi)^3]^{-1}$,
\begin{align}
&\langle\v{k}^{\sigma}\rangle\left(\mbox{at order }\delta^0\right)\equiv\left(\frac{1}{\rho}\frac{2}{(2\pi)^3}\sum_{J=\rm{n,p}}\int_0^{\phi_Jk_{\rm{F}}^J}\v{k}^\sigma n_{\v{k}}^J(\rho,\delta)\d\v{k}
\right)\left(\mbox{at order }\delta^0\right)
=\frac{2[(1-3C_0)\phi_0+3C_0]k_{\rm{F}}^{\sigma+1}}{(\sigma+1)\phi_0}
+\frac{2C_0k_{\rm{F}}^{\sigma+1}(\phi_0^{\sigma-3}-1)}{\sigma-3},\label{ct-0}
\end{align}
and similarly the quadratic and the quartic contributions for $\v{k}^{\sigma}$,
\begin{align}
&\langle
\v{k}^{\sigma}\rangle\left(\mbox{at order }\delta^2\right)\equiv\left(\frac{1}{\rho}\frac{2}{(2\pi)^3}\sum_{J=\rm{n,p}}\int_0^{\phi_Jk_{\rm{F}}^J}\v{k}^\sigma n_{\v{k}}^J(\rho,\delta)\d\v{k}
\right)\left(\mbox{at order }\delta^2\right)\notag\\
=&\frac{3k_{\rm{F}}^\sigma}{\sigma+3}\Bigg[\frac{3C_0\phi_1}{\phi_0}(\phi_1-C_1)+(\sigma+3)C_0C_1\phi_1e^{(\sigma-1)\ln\phi_0}
+\frac{1}{2}\frac{\sigma+3}{\sigma-1}C_0\phi_1^2(\sigma-1)(\sigma-2)e^{(\sigma-1)\ln\phi_0}
+(\sigma+3)C_0C_1\phi_1e^{(\sigma-1)\ln\phi_0}\notag\\
&\hspace*{1.cm}-\left(1+\frac{\sigma}{3}\right)
\Bigg[\frac{3C_0\phi_1}{\phi_0}+3C_0C_1\left(1-\frac{1}{\phi_0}\right)
-(\sigma+3)C_0\phi_1e^{(\sigma-1)\ln\phi_0}
-\frac{\sigma+3}{\sigma-1}C_0C_1\left[e^{(\sigma-1)\ln\phi_0}-1\right]\Bigg]\notag\\
&\hspace*{1cm}+\frac{\sigma(\sigma+3)}{18}\Bigg[1-3C_0\left(1-\frac{1}{\phi_0}\right)+
\frac{\sigma+3}{\sigma-1}C_0\left[e^{(\sigma-1)\ln\phi_0}-1\right]\Bigg]\Bigg],\label{ct-2}\\
&\langle \v{k}^\sigma\rangle\left(\mbox{at order }\delta^4\right)\equiv\left(\frac{1}{\rho}\frac{2}{(2\pi)^3}\sum_{J=\rm{n,p}}\int_0^{\phi_Jk_{\rm{F}}^J}\v{k}^\sigma n_{\v{k}}^J(\rho,\delta)\d\v{k}
\right)\left(\mbox{at order }\delta^4\right)\notag\\
=&\frac{3k_{\rm{F}}^\sigma}{\sigma+3}
\Bigg[\frac{3C_0\phi_1^3}{\phi_0}(\phi_1-C_1)+\frac{\sigma+3}{24}C_0\phi_1^4(\sigma-2)(\sigma-3)(\sigma-4)e^{(\sigma-1)\ln\phi_0}+\frac{\sigma+3}{6}C_0C_1\phi_1^3(\sigma-2)(\sigma-3)e^{(\sigma-1)\ln\phi_0}\notag\\
&\hspace*{1.cm}+\left(1+\frac{\sigma}{3}\right)\Bigg[\frac{3C_0\phi_1^2}{\phi_0}(C_1-\phi_1)
+\frac{\sigma+3}{6}C_0\phi_1^3(\sigma-2)(\sigma-3)e^{(\sigma-1)\ln\phi_0}
+\frac{\sigma+3}{2}C_0C_1\phi_1^2(\sigma-2)e^{(\sigma-1)\ln\phi_0}\Bigg]\notag\\
&\hspace*{1.cm}+\frac{\sigma(\sigma+3)}{18}\Bigg[\frac{3C_0\phi_1}{\phi_0}(\phi_1-C_1)+\frac{\sigma+3}{2}C_0\phi_1^2(\sigma-2)e^{(\sigma-1)\ln\phi_0}
+(\sigma+3)C_0C_1\phi_1e^{(\sigma-1)\ln\phi_0}\Bigg]\notag\\
&\hspace*{1.cm}+\left(\frac{\sigma}{18}-\frac{\sigma^3}{162}\right)
\Bigg[3C_0\left[\frac{\phi_1}{\phi_0}+C_1\left(1-\frac{1}{\phi_0}\right)\right]-(\sigma+3)C_0\phi_1e^{(\sigma-1)\ln\phi_0}-\frac{\sigma+3}{\sigma-1}C_0C_1\left[e^{(\sigma-1)\ln\phi_0}-1\right]\Bigg]\notag\\
&\hspace*{1.cm}+\frac{1}{1944}\sigma(\sigma-6)(\sigma-3)(\sigma+3)\left[1-3C_0\left(1-\frac{1}{\phi_0}\right)
+\frac{\sigma+3}{\sigma-1}C_0\left[e^{(\sigma-1)\ln\phi_0}-1\right]\right]\Bigg].\label{ct-4}
\end{align}
\end{widetext}


\begin{references}

\bibitem{LiBA13} B.A. Li and X. Han, Phys. Lett. \textbf{B727}, 276 (2013).
\bibitem{LiBA19}B.A. Li, P.G. Krastev, D.H. Wen and N.B. Zhang,  Eur. Phys. J. A \textbf{55}, 117 (2019).
\bibitem{LiBA21} B.A. Li, B.J. Cai, W.J. Xie and N.B. Zhang, Universe \textbf{7}, 182 (2021).
\bibitem{Bom91}I. Bombaci and U. Lombardo, Phys. Rev. C \textbf{44}, 1892 (1991).
\bibitem{Lee98} C.H. Lee, T.T.S. Kuo, G.Q. Li, and G.E. Brown, Phys. Rev. C \textbf{57}, 3488 (1998).

\bibitem{Ste06}A.W. Steiner, Phys. Rev. C \textbf{74}, 045808 (2006).
\bibitem{Cai12} B.J. Cai and L.W. Chen, Phys. Rev. C \textbf{85}, 024302 (2012).
\bibitem{Gon17} C. Gonzalez-Boquera, M. Centelles, X. Vinas, and A. Rios, Phys. Rev. C \textbf{96}, 065806 (2017).
\bibitem{PuJ17} J. Pu, Z. Zhang, and L.W. Chen, Phys. Rev. C \textbf{96}, 054311 (2017).


\bibitem{Hen14}O. Hen \textit{et al.}, Science,  \textbf{346}, 614 (2014).
\bibitem{Due18}M. Duer \textit{et al.}, Nature, \textbf{560}, 617 (2018).
\bibitem{Sch19}B. Schmookler \textit{et al.}, Nature, \textbf{566}, 354 (2019).
\bibitem{Sch20}A. Schmidt \textit{et al.}, Nature, \textbf{578}, 540 (2020).
\bibitem{Cai16}B.J. Cai and B.A. Li, Phys. Rev. C \textbf{92}, 011601(R) (2015).

\bibitem{Xulili} C. Xu, A. Li, B.A. Li, Journal of Physics: Conference Series \textbf{420}, 012190 (2013).


\bibitem{Hen15}O. Hen, B.A. Li, W.J. Guo, L. Weinstein, and E. Piasetzky, 
Phys.  Rev.  C \textbf{91}, 025803 (2015).

\bibitem{LiBA15}B.A. Li, W.J. Guo, and Z.Z. Shi, \textbf{91},  044601 (2015).



\bibitem{LCCX18}B.A. Li, B.J. Cai, L.W. Chen, and J. Xu, Prog. Part. Nucl. Phys. \textbf{99}, 29 (2018).

\bibitem{Wel16}C. Wellenhofer, J. Holt, and N. Kaiser, Phys. Rev. C \textbf{93}, 055802 (2016).

\bibitem{And82}T. Ando, A. Fowler, and F. Stern, Rev. Mod. Phys. \textbf{54}, 437 (1982).
\bibitem{Sar11}S. Sarma \textit{et al.}, Rev. Mod. Phys. \textbf{83}, 407 (2011).
\bibitem{Pit16}L. Pitaevskii and S. Strangari, \textit{Bose-Einstein Condensation and Superfluidity}, Part IV, Oxford, 2016.

\bibitem{Blo08}I. Bloch, J. Dalibard, and W. Zwerger, Rev. Mod. Phys. \textbf{80} ,885 (2008).
\bibitem{Lew07}M. Lewenstein \textit{et al.}, Adv. Phys. \textbf{56}, 243 (2007).
\bibitem{Lew12}M. Lewenstein, A. Sanpera, and V. Ahufinger, \textit{Ultracold Atoms in Optical Lattices: Simulating quantum many-body systems}, Oxford, 2012.
\bibitem{Gro17}C. Gross and I. Bloch, Science, \textbf{357}, 995 (2017).


\bibitem{dD1} A. G\"orlitz et al., Phys. Rev. Lett. {\bf 87}, 130402 (2001).
\bibitem{dD2}F. Schreck et al., Phys. Rev. Lett. {\bf 87}, 080403 (2001).
\bibitem{dD3}D. Rychtarik et al., Phys. Rev. Lett. {\bf 92}, 173003 (2004).
\bibitem{dD4}H. Moritz et al., Phys. Rev. Lett. {\bf 94}, 210401 (2005).

\bibitem{Pet16}D.S. Petrov and G.E. Astrakharchik, Phys. Rev. Lett. \textbf{117}, 100401 (2016).
\bibitem{Pet00}D.S. Petrov, M. Holzmann, and G.V. Shlyapnikov, Phys. Rev. Lett. \textbf{84}, 2551 (2000).
\bibitem{Mor15}S. Moroz, J. D'Incao, and D.S. Petrov, Phys. Rev. Lett. \textbf{115}, 180406 (2015).
\bibitem{Ber11}G. Bertaina and S. Giorgini, Phys. Rev. Lett. \textbf{106}, 110403 (2011).

\bibitem{dD5}Y. Nishida and S. Tan,
Phys. Rev. Lett. \textbf{101}, 170401 (2008).

\bibitem{Ber13}G. Bertaina \textit{et al.}, Phys. Rev. Lett. \textbf{110}, 115303 (2013).

\bibitem{Baz18}B. Bazak and D.S. Petrov, Phys. Rev. Lett. \textbf{121}, 263001(2018).


\bibitem{Tan} S. Zhu and S. Tan,
[arXiv:1905.05117 [nucl-th]].

\bibitem{Guo} P. Guo and V. Gasparian, Phys. Rev. D {\bf 97}, 014504 (2018).


\bibitem{Fri05}S. Fritsch,  N. Kaiser, and W. Weise, Nucl. Phys. \textbf{A750}, 259 (2005).

\bibitem{Tod05} B.G. Todd-Rutel and J. Piekarewicz, Phys. Rev. Lett. \textbf{%
95}, 122501 (2005).

\bibitem{Hen16x}O. Hen, G.A. Miller, E. Piasetzky, and L.B. Weistein, Rev. Mod. Phys. \textbf{89}, 045002 (2017).

\bibitem{Carb14}A. Carbone, A. Polls, C. Provid$\hat{\rm{e}}$ncia, A. Rios and I. Vida$\tilde{\rm{n}}$a, Euro. Phys. J. A 50, 13 (2014).

\bibitem{Vid11}I. Vida$\tilde{\rm{n}}$a,  A. Polls, and  C. Provid$\hat{\rm{e}}$ncia, 
Phys.  Rev.  C \textbf{84}, 062801(R) (2011).

\bibitem{Car12}A. Carbone, A. Polls, and A. Rios, 
Eur.  Phys.  Lett. \textbf{97}, 22001 (2012).
\bibitem{Lov11}A. Lovato, O. Benhar, S. Fantoni,  A. Illarionov, and K. Schmidt, 
Phys.  Rev.  C \textbf{83}, 054003 (2011).
\bibitem{Cai16a}B.J. Cai and B.A. Li, Phys. Rev. C \textbf{93},  014619 (2016).
\bibitem{Sou20-a}L.A. Souza \textit{et al.}, arXiv: 2004.10309 (2020).
\bibitem{Sou20-b}L.A. Souza \textit{et al.}, Phys. Rev. C \textbf{101}, 065202 (2020).
\bibitem{HLu22}H. Lu, Z.Z. Ren, and D. Bai, Nucl. Phys. \textbf{A1021}, 122408 (2022).

\bibitem{Lou22}
O. Louren\c{c}o, T. Frederico, and M. Dutra,
Phys. Rev. D \textbf{105}, 023008 (2022).

\bibitem{Ant93}A.N. Antonov, P.E. Hodgson, and I.Zh. Petkov, Nucleon Correlations in Nuclei, Springer-Verlag, Berlin, Heidelberg 1993, ISBN 3-540-55911-6.

\bibitem{Yong17-a}G.C. Yong and B.A. Li, Phys. Rev. C \textbf{96}, 064614 (2017).
\bibitem{Yong17-b}G.C. Yong, Phys. Lett. \textbf{776}, 447 (2017).
\bibitem{Wang17}Z. Wang \textit{et al.}, Phys. Rev. C \textbf{96},  054603 (2017).
\bibitem{Guo21}W.M. Guo, B.A. Li, and G.C. Yong, Phys. Rev. C \textbf{104},  034603 (2021).

\bibitem{Chris}
K. Hagel and J.B. Natowitz,
[arXiv:2111.09399 [nucl-ex]].

\end{references}
\end{document}